\documentclass[11pt]{article}

\usepackage{amssymb,amsmath,latexsym,bm,amsfonts}
\usepackage{slashed}
\usepackage{graphicx}
\usepackage{longtable}
\usepackage{color,xcolor}
\usepackage{indentfirst}
\usepackage{subfigure}
\usepackage[margin=1in]{geometry}
\usepackage{mathtools,tensor}
\usepackage{footmisc}

\usepackage{breqn}
\usepackage{authblk}
\usepackage{etoolbox}
\numberwithin{equation}{section}
\usepackage{stmaryrd}

\usepackage[nottoc ]{tocbibind}

\newcommand{\be}{\begin{equation}}
\newcommand{\ee}{\end{equation}}
\newcommand{\beqn}{\begin{eqnarray}}
\newcommand{\eeqn}{\end{eqnarray}}

\newcommand{\ta}{\theta}
\newcommand{\taa}{\theta\theta}

\newcommand{\tab}{\bar\theta }
\newcommand{\taab}{\bar\theta\bar\theta}

\newcommand{\ldbl}{ \lambda \slashed{\partial}\bar{\lambda}}
\newcommand{\dlbl}{\partial_{\mu}\lambda \sigma^{\mu}\bar{\lambda}}
\newcommand{\lbdbl}{\ldb \bar{\slashed \p}\ld}
\newcommand{\lsl}{\lambda \sigma^{\mu} \bar{\lambda}}
\newcommand{\cA}{\mathcal{A}}
\newcommand{\cB}{\mathcal{B}}
\newcommand{\cE}{\mathcal{E}}
\newcommand{\cM}{\mathcal{M}}
\newcommand{\ldb}{\bar{\lambda}}
\newcommand{\ld}{\lambda}
\newcommand{\Db}{\bar{\sfD}}
\newcommand{\sfD}{\textsf D}

\newcommand{\G}{Z}
\newcommand{\sG}{\sqrt{Z}}
\newcommand{\slp}{\slashed{\partial}}

\newcommand{\sGo}{\sqrt{Z_0}}
\newcommand{\M}{\mathcal{M}}

\newcommand{\p}{\partial}
\newcommand{\Ft}{\tilde{F}}
\newcommand{\kp}{\kappa^2}
\newcommand{\kpp}{\kappa^4}

\DeclareMathOperator{\Imag}{Im}

\usepackage{breqn}
\usepackage[colorlinks=true,filecolor=black,citecolor=blue ]{hyperref}
\usepackage{cleveref}

\usepackage{eufrak}
 \usepackage{mathrsfs}
 \usepackage{mathrsfs}

\begin{document}

\renewcommand{\baselinestretch}{1.2}
\setlength{\parindent}{2em}
\setlength{\parskip}{0.4em}

\begin{titlepage}
  \ \\
  \vskip 1cm

  \begin{center}

    \huge{Note on supersymmetric Dirac-Born-Infeld action \\ with Fayet-Iliopoulos term}

    \vspace{0.8 cm}
    \normalsize{Ignatios Antoniadis~$^{a,b}$, Hongliang   Jiang~$^{b}$ and Osmin Lacombe~$^{a}$}
    
    \vspace{0.5cm}
    \normalsize{\it  $^{a}$Laboratoire de Physique Th\'eorique et Hautes Energies - LPTHE \\
Sorbonne Universit\'e, CNRS, 4 Place Jussieu, 75005 Paris, France \\ [10pt]
$^{b}$Albert Einstein Center, Institute for Theoretical Physics, University of Bern, \\
 			Sidlerstrasse 5, 3012 Bern, Switzerland } \\
    \vspace{0.5 cm}

{\texttt{antoniadis@itp.unibe.ch}}, {\texttt{jiang@itp.unibe.ch}}, {\texttt{osmin@lpthe.jussieu.fr}}

    \vspace{0.8 cm}

    \begin{abstract}
We study the  Dirac-Born-Infeld (DBI) action with one linear and one non-linear supersymmetry in the presence of a constant Fayet-Iliopoulos (FI) D-term added explicitly or through a deformation of supersymmetry transformations. The linear supersymmetry appears to be spontaneously broken since the D auxiliary field  gets a non-vanishing vacuum expectation value and an extra term proportional to the FI parameter involving fermions emerges in the non-linear formulation of the action written recently. However in this note, we show that on-shell this action is equivalent to a standard supersymmetric DBI action {\it without} FI term but with redefined tension, at least up to order of mass-dimension 12 effective interactions.

 \end{abstract}

    \vspace{1cm}
      
  \end{center}
  
\end{titlepage}

\pagestyle{plain}
\pagenumbering{arabic}

 {\hypersetup{linkcolor=black}
 \tableofcontents
}

\newpage

\section{Introduction}\label{Intro}The  four dimensional supersymmetric Dirac-Born-Infeld (SDBI) action describes the effective field theory of a D3-brane and breaks half of the supersymmetries of the bulk, which are non-linearly realized on the D-brane world-volume. In the physically interesting case the bulk has $\mathcal{N}=2$ supersymmetry, obtained for instance upon compactification of the ten-dimensional type II superstrings on a Calabi-Yau threefold. The goldstino of the $\mathcal{N}=2\to \mathcal{N}=1$ partial breaking belongs to a vector multiplet of the linear supersymmetry that has non-trivial self-interactions due to the non-linear supersymmetry. The SDBI action can be obtained from the $\mathcal{N}=2$ quadratic action by imposing a non-linear constraint that eliminates the $\mathcal{N}=1$ chiral superfield in terms of the $U(1)$ vector-goldstino superfield that form together the $\mathcal{N}=2$ vector multiplet~\cite{Bagger:1996wp, Rocek:1997hi}. This constraint takes a very simple nilpotent form in terms of the deformed $\mathcal{N}=2$ vector superfield which incorporates a deformation in the transformations of one (the would be non-linear) supersymmetry~\cite{Antoniadis:2008uk, Antoniadis:2017jsk, Antoniadis:2019gbd}. Note that the deformation cannot be interpreted as an expectation value of an auxiliary field. The resulting action depends on one parameter, the goldstino decay constant $\kappa$, or equivalently the D3-brane tension.

After solving the constraint, the SDBI action can be written as an integral over the $\mathcal{N}=1$ superspace of an expression involving the gauge-field strength superfield where non-linear supersymmetry is not manifest~\cite{Antoniadis:2008uk}. In terms of components, upon superspace integration, the bosonic part of the action can be written in a closed form, while it is not the case for the part involving the $U(1)$ gauginos. On the other hand, an explicit form of the whole action was given within the formalism of non-linear supersymmetry, using variables where manifest linear supersymmetry is lost~\cite{Bellucci:2015qpa, Cribiori:2018dlc}. The two actions are of course equivalent on-shell, related through field redefinitions. 

The goal of this note is to study the uniqueness of the SDBI action which is an interesting open problem that has not yet been settled. Indeed, in the case of  $\mathcal{N}=1$ non-linear supersymmetry, the Volkov-Akulov (VA) action~\cite{Volkov:1972jx} was shown to be unique up to field redefinitions that do not change the theory on-shell (see e.g.~\cite{Kuzenko:2010ef}). In this work, we address this problem for the SDBI action within effective field theory, following a similar strategy. Of course, a natural arena of studying this action and its uniqueness is within string theory. 

A comparison   of  SDBI with the corresponding amplitudes in string theory was done in~\cite{Tseytlin:1999dj,Bergshoeff:2001dc} at the level of mass dimension-8 interactions involving four bosons, four fermions, two bosons and two fermions, i.e. $F^4$, $(\lambda\partial\ldb)^2$ and $(\lambda\partial\ldb)F^2$, schematically. Actually, dimension-6 interactions involving two fermions and one boson also appear in the form $(\lambda\partial\ldb)F$.  Nevertheless, these terms vanish on-shell and can be eliminated by field redefinitions, giving rise to extra contributions in dimension 8. It turns out however that dimension-8 operators involving fermions are completely determined by $\mathcal{N}=1$ non-linear supersymmetry alone and do not provide any non-trivial test of string theory. Indeed, the coefficient of $(\lambda\partial\ldb)^2$ is fixed by the standard $\mathcal{N}=1$ non-linear VA action, while $(\lambda\partial\ldb)F^2$ is uniquely fixed by the universal goldstino coupling through the energy-momentum tensor, using the same decay constant~\cite{Antoniadis:2004uk}, dictated by the low-energy theorem of spontaneous supersymmetry breaking~\cite{Samuel:1982uh, wess1992supersymmetry}. On the other hand, the bosonic $F^4$ operator follows from the expansion of the DBI action which is completely determined from the combination of linear and non-linear supersymmetries~\cite{Bagger:1996wp, Rocek:1997hi}. 
It is therefore necessary to go beyond dimension-8 operators in order to make a non-trivial test of the fermionic dependence of the D-brane action and its comparison with the SDBI form.

 In order to get a better understanding of the uniqueness of the SDBI action, one can consider deformations that do not change the form of the action bosonic part. This can be achieved by adding
Fayet-Iliopoulos (FI) D-terms and their generalisations. The generation mechanisms and the effect of such terms constitute an interesting open problem  related to supersymmetry breaking, that   becomes  more restrictive in supergravity and even more in extended supersymmetric theories. New FI terms, that do not require gauging the R-symmetry when coupled to supergravity, were written recently within $\mathcal{N}=1$~\cite{Cribiori:2017laj, Antoniadis:2018cpq,Antoniadis:2018oeh,Antoniadis:2019nwz} and $\mathcal{N}=2$~\cite{Antoniadis:2019hbu} supersymmetry; their bosonic part is identical to an ordinary constant FI term, while their fermionic dependence is highly non-trivial. 
The first step of these studies is obviously the analysis of the effects of the standard FI term added to the SDBI action, that we discuss in this work.

In the absence of matter, a constant FI term can be added in the SDBI action since it is invariant under both supersymmetries. Its presence drives the $\sfD$ auxiliary field of the $U(1)$ vector multiplet to get a non-vanishing vacuum expectation value (VEV), breaking spontaneously apparently the linear supersymmetry. However, after elimination of the auxiliary field, the bosonic part of the action acquires again the SDBI form by redefining $\kappa$, and thus the D-brane tension~\cite{Antoniadis:2008uk, Kuzenko:2009ym}. Nevertheless possible differences could come from fermionic contributions. Indeed, it was recently argued  that in the non-linear formulation where fermion dependence can be written explicitly, there is an extra term proportional to the FI parameter~\cite{Cribiori:2018dlc}. This term involves an odd number of bosons, in contrast to the standard SDBI action which involves only even numbers, upon elimination of the auxiliary field. It is therefore interesting to study the effect of the new term on physical amplitudes and eventually compare them with corresponding string amplitudes. Also, more recently, another kind of an FI D-term was constructed by considering the most general $\mathcal{N}=2$ deformation and imposing the nilpotent constraint on the corresponding $\mathcal{N}=2$ vector superfield, in the presence of a non-vanishing $\theta$ angle in the action~\cite{Antoniadis:2019gbd}. The advantage of this mechanism is that a constant FI D-term is not added by hand but is induced from an action which is manifestly gauge invariant (and not just up to a total derivative, which usually makes it difficult to couple it to supergravity). The relevant deformation amounts to add an imaginary part $\gamma$ to the $\sfD$ auxiliary field.


In this note we compute for the first time non-trivial higher dimensional physical vertices 
corresponding to fermionic contributions in the SDBI action, in the presence of a standard or induced FI D-term. As mentioned above, such terms being linear in the $\mathcal{N}=1$ vector multiplet, they generate interactions containing an odd number of gauge fields. As these terms are not present in the standard SDBI action, we expect that they  show the difference between the SDBI action and SDBI coupled to FI terms in appropriate on-shell physical operators. 

The first possible appearance of an odd number of gauge fields is at dimension-6 level, through terms containing two gauginos and one field strength. Nevertheless, as we mentioned above, these terms are not physical and can be eliminated by means of field redefinitions. The dimension-8 terms are either fully bosonic with four field strengths, fully fermionic with four gauginos, or mixed with two gauginos and two gauge field strengths. The fully bosonic term follows from the standard bosonic part of the SDBI action with a redefined tension that takes into account the coefficient of the FI D-term (or the $\gamma$-deformation and $\theta$-angle in the induced case). On the other hand, the fully fermionic term follows from the Volkov-Akulov action describing the dynamics of the gaugino field, which is the goldstino of partial supersymmetry breaking, with a decay constant fixed by the brane tension due to the linear supersymmetry. Eventually, the only non-trivial term is the mixed gaugino -- gauge field strength one, which is however completely determined by the standard goldstino coupling to energy-momentum tensor, fixed by the low-energy theorems of supersymmetry breaking. There is no other dimension-8 operator involving gauge fields coupled to the goldstino, unlike other operators involving scalars or extra fermions~\cite{Antoniadis:2004uk}.


 The above considerations motivate our present work to compute dimension-10 physical mixed operators involving gauginos and gauge fields -- either four gauginos and two gauge bosons, or two gauginos and three gauge fields -- in both cases of the SDBI action with a standard or induced FI term, referred in the following to as SDBI+FI or SDBI+$\gamma,\theta$, respectively. In the first case of SDBI+FI, we use both the non-linear formalism and the standard constrained superfield one with manifest   linear supersymmetry, while in the second case of SDBI+$\gamma,\theta$ we use only the constrained superfield method. We find that all dimension-10 operators can be eliminated by field redefinitions in both cases, strongly suggesting that the presence of FI terms   does  not break the linear supersymmetry but just   modifies  the goldstino decay constant. Actually, in the SDBI+FI case, using the formalism of non-linear supersymmetry, we also compute a class of dimension-12 mixed operators with two gauginos and four bosons, of the form $(\ld\p\ldb) F^4$, and show further support to the above statement. In the context of string theory, this implies that FI terms rotate the D3-brane, and thus the directions of linear and non-linear supersymmetry in the bulk without breaking them~\cite{Kuzenko:2009ym, Antoniadis:2019gbd}. Breaking of supersymmetry is expected to occur when an extra reference brane is inserted without FI term, or in the presence of charged matter like in intersecting brane configurations~\cite{Antoniadis:2019gbd}.

Let us stress that although the final result has a simple interpretation within string theory, it was not apriori obvious without explicit computation. Moreover, the possibility of deforming the SDBI action by new FI D-terms remains open, as well as their possible generation in string theory. Such terms are expected to modify (on-shell) the fermionic dependence of higher than dimension-8 operators, motivating corresponding computations in string theory.
 
 Our paper is organized as follows. In \cref{sec2NL} we  start with the non-linear supersymmetry representation of the SDBI+FI action and compute the physical on-shell action by means of  field redefinitions. Especially, we show  the absence of the dimension-6 and dimension-10 terms, and compute the dimension-12 terms of the form $(\ld\p\ldb)F^4$. 
  In \cref{sectionexpansion} we  consider the  SDBI+$\gamma,\theta$ action in the formalism with manifest linear supersymmetry and use  field redefinitions to compute its physical on-shell action up to dimension 10. In a specific limit, we also obtain the on-shell action of SDBI+FI which agrees with the one obtained from non-linear supersymmetry formalism. 
  We   conclude in \cref{Summary}.   To simplify presentation of our technical analysis, we include two appendices. \Cref{usefulids} contains our conventions and some useful identities, and \Cref{somedetails} contains some technical details including relevant superfield expansions.


   
\section{SDBI action with standard FI term}  \label{sec2NL} 

The goal of this section is to compute the physical on-shell Lagrangian of the  SDBI model with  a standard  Fayet-Iliopoulos term, which will be referred to as SDBI+FI Lagrangian in the following. The SDBI action can be constructed by deforming an $\mathcal N=2$ vector multiplet and imposing nilpotent constraints, as we will briefly review in the next section.  The resulting SDBI action then possesses a manifest $\mathcal N=1$ linear supersymmetry as well as a hidden non-linear $\mathcal N=1$  supersymmetry. 

As a starting point, here, we will use the SDBI+FI action expressed in  the non-linear formalism, where the complete action can be written explicitly in terms of variables appropriate for the non-linear supersymmetry but at the cost of losing manifest invariance under linear supersymmetry~\cite{Cribiori:2018dlc}. We will first expand it in components in \cref{subsecexpansionfromNL} and  then     use field redefinitions to spell out  the on-shell physical interactions with dimension  less than 12, as well as all dimension-12 operators of the form $(\lambda\p\bar\lambda) F^4$ in section~\ref{subsecphysicalNL}.
The result suggests that  on-shell the FI term plays a trivial role and can be absorbed by redefining the brane tension. 

\subsection{SDBI+FI from non-linear supersymmetry formalism}\label{subsecexpansionfromNL}
 
 The non-linear supersymmetry formalism was systematically developed in~\cite{Cribiori:2018dlc}. Applying the formalism to the SDBI+FI model, the   Lagrangian can be expressed as\,\footnote{ 
The two terms in the parenthesis have the same sign, in agreement with~\cite{Bellucci:2015qpa} but in contrast to the opposite sign in~\cite{Cribiori:2018dlc}. 
The coupling constants are related to those of \cite{Cribiori:2018dlc} as follows: $\kappa = 1/({2\sqrt{2}m})$ and $ {g^2} ={m}/\beta$. The dual tensor $\tilde{\mathcal{F}}$ is defined as in \eqref{dualtensors} with a factor of $i$. 
\label{footnote1}
}
\beqn \label{Lnonlin}
\mathcal{L}_{\rm SDBI+FI} &=& - \frac1{8\kappa^2g^2} \det{\bm A} \left(1+\sqrt{1+16g^4\kappa^2\xi^2}\sqrt{1+ 4\kappa^2 \mathcal{F}^2+4\kappa^4 (\mathcal{F}\tilde{\mathcal{F}})^2}\right)\nonumber \\
&& +2\sqrt{2}i \,\kp\xi \det{\bm A} \epsilon^{abcd}[({\bm A}^{-1})_{a}^{~\nu}\partial_{\nu} \lambda]\sigma_b[({\bm A}^{-1})_{c}^{~\rho}\partial_{\rho}\bar{\lambda}]({\bm A}^{-1})_d^{~\mu}u_{\mu} ~,
\eeqn
 where   $\lambda$  is the goldstino in the non-linear superymmetry formalism which is also the gaugino in the linear one through field redefinition
 \footnote{We use the same symbol $\lambda$ to denote both the  goldstino in this section   and the gaugino in the   section~\ref{sectionexpansion} for simplicity of notation.}; 
 $u_{\mu}$ is the gauge boson; and the composite gauge field strength $\mathcal F_{ab}$ is defined by
 \be
\mathcal F_{ab}=({\bm A}^{-1})^{\mu}_a ({\bm A}^{-1})^{\nu}_b F_{\mu \nu}~, \qquad F_{\mu \nu}=\p_{\mu} u_{\nu} -\p_{\nu} u_{\mu}~,
 \qquad \mathcal F^2 \equiv  \mathcal F_{ab} \mathcal F^{ab}~,\qquad  \mathcal F \tilde{\mathcal F}\equiv \mathcal F_{ab} \mathcal {\Ft}^{ab}~,
\ee 
 where ${\bm A}$ is the vielbein matrix   
 \beqn \label{VA}
{\bm A}^a_{\mu}= \delta^a_{\mu} +  i \kappa^2 \ld \sigma^a \partial_{\mu} \bar{\ld} -  i \kappa^2 \partial_{\mu} \ld \sigma^a \bar{\ld} ~.    
\eeqn
As shown in appendix \ref{derivationOfgaugeInv},  the Lagrangian \eqref{Lnonlin} can be further  rewritten in a manifestly gauge invariant way as
 \beqn\label{Lnonlintsf}
 \mathcal{L}_{\rm SDBI+FI} &=& - \frac1{8\kappa^2g^2} \det{\bm A} \left(1+\sqrt{1+16g^4\kappa^2\xi^2}\sqrt{1+ 4\kappa^2 \mathcal{F}^2+4\kappa^4 (\mathcal{F}\tilde{\mathcal{F}})^2}\right)\nonumber \\
&& -2\sqrt{2}\,\kp\xi \,\lambda \sigma_{\gamma} \partial_{\rho}\bar{\lambda} ~ \Ft^{ \gamma \rho} ~.  
\eeqn

We are going now to expand the Lagrangian in operators of increasing mass dimension. Let us first recall the mass dimensions of various fields and couplings 
\be
[u]=1\,, \qquad [F]=2\,, \qquad [\lambda]=\frac32\, ,\qquad  [\kappa]=-2\, , \qquad [g]=0\, , \qquad [\xi ]= 2\,. 
\ee
We then expand in components each part of this Lagrangian, up to terms of dimension 12 -- except for  dimension-12 terms of the form $(\lambda\p\bar\lambda)^2 F^2$  which are more involved. In the following, all equalities are written up to total derivatives, or terms with mass  dimension higher than   12,  or $(\lambda\p\bar\lambda)^2 F^2$ terms. 

For spinors, we  follow the  conventions in  \cite{wess1992supersymmetry}. Some useful relations used for the computations can be found in appendix~\ref{usefulids}. 

 The $\det{\bm A}$ part of~\eqref{Lnonlin} is related to the Volkov-Akulov action \cite{Volkov:1972jx} describing goldstino dynamics 
 \beqn \label{VALagrangianbeforefieldredef}
 {\cal L}_{\rm VA} &=&- \frac{1}{2 \kappa^2} \det{\bm A} \nonumber \\
 &=&-\frac{1}{2 \kappa^2} \left( 1+ \kappa^2 i \ld \slashed{\partial} \bar{\ld} + \kappa^2 i\ldb \bar{\slashed \p}\ld - 2 \kappa^4 \left( (\ld \slashed{\partial} \bar{\ld})(\ldb \bar{\slashed \p}\ld) + \ld \sigma^{\mu}\partial_{\nu} \bar{\ld} \partial_{\mu} \ld \sigma^{\nu} \bar{\ld } \right) \right) + O(\ld^6) \nonumber \\
 &=&-\frac{1}{2 \kappa^2} -\frac i2 \ld \slashed{\partial} \bar{\ld} -\frac i2 \ldb \bar{\slashed \p}\ld -\frac {\kappa^2}2 \left(- (\ld \slashed{\partial} \bar{\ld})( \ldb \bar{\slashed \p}\ld) - 2 ( \ld \slashed{\partial} \bar{\ld})^2 - 2 (\ldb \bar{\slashed \p}\ld)^2 + \partial_{\mu}(\ld^2)\partial^{\mu}(\bar{\ld}^2)\right. \nonumber \\[5pt]
 &&  +  \left. \bar{\ld}^2 \ld \Box \ld + \ld^2 \bar{\ld} \Box \bar{\ld} \vphantom{{( \ld \slashed{\partial} \bar{\ld})^2}^2} \right) + O(\ld^6) + \text{total derivatives}  \,.
 \eeqn
Here and in the rest of the paper, we use  the following conventions
\be 
 \lambda \slashed \p \bar \chi\equiv\lambda  \sigma^\mu \p_\mu  \bar \chi ,\qquad
   \bar \lambda \bar\sigma^\mu \p_\mu \chi\equiv \bar \lambda  \bar { \slashed  \p   }  \chi, \qquad
 (\lambda \slashed \p \bar \chi)^*= -\bar \lambda  \bar { \slashed  \p   }  \chi~.
\ee
When expanding $\mathcal{F}^2$ in \eqref{Lnonlin}, one finds the gauge boson kinetic term and higher order interactions
\beqn \label{F2terms}
 \mathcal{F}^2 &=& \eta^{ac} \eta^{bd} \mathcal F_{ab} \mathcal F_{cd} = \eta^{ac} \eta^{bd} ({\bm A}^{-1})^{\mu}_a ({\bm A}^{-1})^{\nu}_b ({\bm A}^{-1})^{\rho}_c ({\bm A}^{-1})^{\sigma}_d F_{\mu\nu} F_{\rho \sigma} \nonumber \\
 &=& \eta^{ac} \eta^{bd} F_{\mu\nu} F_{\rho \sigma} \delta^{\mu}_a \delta^{\nu}_b\delta^{\rho}_c\delta^{\sigma}_d +  4\kp \eta^{ac} \eta^{bd} F_{\mu\nu} F_{\rho \sigma} (- i \lambda \sigma^{\mu} \partial_{a} \bar{\lambda} + i \partial_{a} \lambda \sigma^{\mu} \bar{\lambda})  \delta^{\nu}_b\delta^{\rho}_c\delta^{\sigma}_d + \cdots \nonumber \\
 &=& F^2 +  4\kp( i \lambda \sigma^{\mu} \partial_{\rho} \bar{\lambda} - i \partial_{\rho} \lambda \sigma^{\mu} \bar{\lambda})  F_{\mu\nu} F^{\nu \rho } + \cdots.
\eeqn
The same computation for $(\mathcal{F}\tilde{\mathcal{F}})^2$ gives
\beqn\label{FFdualsquare}
(\mathcal{F}\tilde{\mathcal{F}})^2&=&\left( F\Ft +  4\kp( i \lambda \sigma^{\mu} \partial_{\rho} \bar{\lambda} - i \partial_{\rho} \lambda \sigma^{\mu} \bar{\lambda})  F_{\mu\nu} \Ft^{\nu \rho } + O((\lambda\p\bar\lambda)^2 F^2)\right)^2\nonumber\\ 
&=&\left(F\Ft - \kp(i\ldbl+i\lbdbl)F \Ft+O((\lambda\p\bar\lambda)^2 F^2)\right)^2 \nonumber\\
&=&(F\Ft)^2- 2 \kp(i\ldbl+i\lbdbl) (F\Ft)^2+O((\lambda\p\bar\lambda)^2 F^2\big)\,. 
\eeqn
The same expansion is also obtained directly by noticing that $\mathcal{F}\tilde{\mathcal{F}}=   \det{\bm A}^{-1} F\Ft $.

Collecting all the above terms, the SDBI+FI Lagrangian becomes
\beqn\label{Lagrangianfordim12NL}
 \mathcal{L}_{\rm SDBI+FI}
 &=& A \left[ \vphantom{A^{a'}}1+\kp i \ldbl +  \kp i \lbdbl - \kpp (\ld \slashed{\partial} \bar{\ld}) (\lbdbl) - 2 \kpp ( \ld \slashed{\partial} \bar{\ld})^2 - 2 \kpp (\lbdbl)^2 \right. \nonumber\\[2pt]
 &&  \left. \quad+ \kpp \bar{\ld}^2 \ld \Box \ld + \kpp \ld^2 \bar{\ld} \Box \bar{\ld} +\kpp \partial_{\mu}(\ld^2)\partial^{\mu}(\bar{\ld}^2) \vphantom{A^{a'}} \right] +2{B}\kpp(i\ldbl+i\lbdbl)F^2 \nonumber \\[2pt]
&&+8{B}{\kappa^4}\left ( i \lambda \sigma^{\mu} \partial_{\rho} \bar{\lambda} - i \partial_{\rho} \lambda \sigma^{\mu} \bar{\lambda}\right)  F_{\mu\nu} F^{\nu \rho }-16{B}{\kappa^6}\left ( i \lambda \sigma^{\mu} \partial_{\rho} \bar{\lambda} - i \partial_{\rho} \lambda \sigma^{\mu} \bar{\lambda}\right)  F_{\mu\nu} F^{\nu \rho }F^2 \nonumber \\[2pt]
&& +2 B\kp \left({F^2}+\kp (F\Ft)^2-\kp F^4\right) + C \kp\lambda \sigma^{\mu} \partial^{\nu}\bar{\lambda} ~ \Ft_{ \mu \nu} + \cdots~.
\eeqn
The constants $A,B$ and $C$ are defined as
\beqn \label{defABC}
A&\equiv& - \frac{1}{8\kappa^2g^2} \left(1+\sqrt{1+16g^4\kappa^2\xi^2}\right) , \nonumber \\
B &\equiv&-\frac{1}{8\kappa^2g^2} \sqrt{1+16g^4\kappa^2\xi^2}\, ,  \\
C &\equiv& -2\sqrt{2} \xi \,,  \nonumber
\eeqn
and have   mass dimensions $  [A]=[B]=4$ and  $[C]=2 $.

\subsection{Physical action with standard FI term}\label{subsecphysicalNL}

Our goal is to get the physical on-shell Lagrangian. The main strategy is to use field redefinitions to eliminate various on-shell vanishing unphysical   terms and get the physical on-shell  higher dimensional operators. The S-matrix is invariant under field redefinitions.  Thus, to eliminate an unphysical  term of certain dimension, we use a specific field redefinition and act it on terms of lower dimension. However, the field redefinition also acts on other terms in the Lagrangian giving rise to many extra higher dimensional terms. Repeating this procedure allows us to eliminate all unphysical terms and get the on-shell Lagrangian. In general, this process is complicated and tedious. 

Before performing the computations, it is worth pointing out a big simplification. At any step, we will only be interested in the physical Lagrangian up to some dimension, say dimension $\ell$, and thus will always neglect  terms with dimension 
higher than $\ell$.  The simplification occurs if the term $\mathcal O$ under consideration is proportional to an equation of motion of the free theory. In such a  case,  we must be able to  eliminate $\mathcal O$ through certain field redefinition acting on the free kinetic terms. If the dimension of $\mathcal O$ is close to $\ell$, acting the field redefinition on other terms of the Lagrangian may only generate terms with dimension strictly higher than $\ell$. If this is indeed the case, we do not need to work out the field redefinition explicitly  and  can  simply discard the term $\mathcal O$. This circumstance brings us a big simplification.

 To obtain the physical SDBI+FI action we proceed as follows: we first eliminate the lowest dimensional non-physical operators, namely the dimension-6 ones, by means of field redefinitions acting on kinetic terms. We then compute the higher dimensional contributions coming from the field redefinitions acting on the other   terms in the Lagrangian. We repeat this procedure for   operators with higher and higher dimensions.

In the computations, we will make full use of the identities given in \cref{usefulids}. In all equalities thereafter, ellipses $``\cdots"$ should  be understood as total derivatives or higher dimensional terms which we are not interested in. 

\paragraph{Field redefinition \raisebox{.5pt}{\textcircled{\raisebox{-.9pt} {1}}} } To eliminate the dimension-6 term contained in the last line of \eqref{Lagrangianfordim12NL} we apply the following field redefinition
\be \label{fieldredfandb}
\raisebox{.5pt}{\textcircled{\raisebox{-.9pt} {1}}} \quad \ld_{\alpha}\rightarrow \lambda_{\alpha}+ ia (\sigma^{\mu \nu} \lambda )_{\alpha} F_{\mu \nu}, \qquad a=\frac{C}{4A} \in \mathbb{R}, \quad [a]=-2.
\ee
Note that due to the equality  $\sigma^{\rho\gamma}=\frac i2\epsilon^{\rho\gamma\mu\nu} \sigma_{\mu\nu}$, this field transformation is equivalent to the one with $F_{\mu\nu}$ replaced by $\Ft_{\mu\nu}$. The fermion kinetic terms transform as
\beqn\label{gauginotransfieldredef1}
 i\ldbl +{\rm c.c.}&\underset{\eqref{fieldredfandb}}{\longrightarrow}& i\ldbl +{\rm c.c.} - 4a \lambda \sigma^{\mu}\partial^{\nu} \bar{\lambda} \Ft_{\mu \nu} - 2a \lsl \partial^{\nu} F_{\nu \mu} +2  {a}^2 i \lsl \Ft_{\nu \mu} \p_{\rho}F^{\rho \nu}\nonumber \\[-12pt]
&&+ \left( 2 {a}^2  i \lambda \sigma^{\mu} \partial^{\nu} \bar{\lambda} F_{\mu\rho} F^{\rho}_{~ \nu } + \frac{{a}^2}2 i\ldbl F^2 + {\rm c.c.} \right) + \text{total derivatives}~,
\eeqn
and they indeed cancel the dimension-6 terms of \eqref{Lagrangianfordim12NL} with the chosen parameter $a$. The dimension-6 term itself transforms as
\beqn
\ld\sigma^{\mu}\partial^{\nu}\ldb \Ft_{\mu\nu}&\underset{\eqref{fieldredfandb}}{\longrightarrow}& \frac 12 \ld\sigma^{\mu}\partial^{\nu}\ldb \Ft_{\mu\nu}-i a\lambda \sigma^{\mu}\partial^{\nu} \bar{\lambda} F_{\mu\rho} F^{\rho}_{~\nu} -\frac{{a}}2 i\ldbl F^2  -\frac{{a}}4i\ldbl F\Ft\nonumber  \\[-2pt]
&&\hspace{-0.2cm}+\frac{{a}^2}4  \lambda \sigma^{\mu} \partial^{\nu} \bar{\lambda} \left( \Ft_{\mu\nu} F^2 -F_{\mu\nu}F\Ft +\Ft_{\mu\nu} F\Ft-F_{\mu\nu}F^2\right)+{ \rm c.c.} +\cdots~. \hspace{0.5cm}
\eeqn
Other terms in the Lagrangian transform as
\beqn
&& i \lambda \sigma^{\mu} \partial_{\nu} \bar{\lambda} F_{\mu\rho} F^{\rho \nu } + {\rm c.c.} \underset{\eqref{fieldredfandb}}{\longrightarrow}  i \lambda \sigma^{\mu} \partial_{\nu} \bar{\lambda}  F_{\mu\rho} F^{\rho \nu }  -{a} \lsl F_{\alpha \nu} F^{\nu\rho} \partial_{\rho} F^{\alpha}_{~\mu}  +\frac{a}2 \lambda \sigma^{\mu} \partial^{\nu} \bar{\lambda}  F_{\mu\nu} F\Ft\nonumber \\[-10pt]
&&\hspace{4.5cm} -i\frac{{a}^2}4 \lsl F^2 \partial_{\mu}(F\Ft) -\frac{{a}^2}2 i \lambda \sigma^{\mu} \partial_{\nu} \bar{\lambda} F_{\mu\rho} F^{\rho \nu }F^2 +{\rm c.c.} + \cdots~, \hspace{1cm}\\[5pt]
\label{dim10fromdetA}
&& \bar{\ld}^2 \ld \Box \ld + {\rm c.c.}   \underset{\eqref{fieldredfandb}}{\longrightarrow}  \bar{\ld}^2 \ld \Box \ld  + 2ia \ldb^2 \ld \sigma^{\mu\nu} \p_{\rho} \ld ~ \p^{\rho} F_{\mu \nu} + {\rm c.c.} + \cdots  ~.
  \eeqn
\paragraph{Field redefinition \raisebox{.5pt}{\textcircled{\raisebox{-.9pt} {2}}} } 
Although the field redefinition \eqref{Lagrangianfordim12NL}  eliminates the original  dimension-6 term in \eqref{Lagrangianfordim12NL}, it introduces another dimension-6 operator in~\eqref{gauginotransfieldredef1}.  
 Hence, we must combine the field redefinition \eqref{Lagrangianfordim12NL}  with another field redefinition on the gauge boson
\beqn \label{fielddefbosonstrange}
\raisebox{.5pt}{\textcircled{\raisebox{-.9pt} {2}}}\quad u_{\mu}\rightarrow u_{\mu}+b \lambda \sigma_{\mu} \bar{\lambda}~ , \qquad b=-\frac{C}{16B} \in \mathbb{R}, \qquad [b]=-2~.
\eeqn
This is equivalent to the following field-strength redefinition
\beqn \label{transstrength}
F_{\mu \nu}\rightarrow F_{\mu \nu}+b \partial_{\mu}(\lambda \sigma_{\nu} \bar{\lambda}) - b \partial_{\nu}(\lambda \sigma_{\mu} \bar{\lambda}) \equiv F_{\mu \nu}+ 2b  \partial_{[\mu}(\lambda \sigma_{\nu]} \bar{\lambda}) ~.
\eeqn
The gauge boson kinetic term transforms as
\beqn \label{transbosonkinetic}
 {F} ^2 &\underset{\eqref{fielddefbosonstrange}}{\longrightarrow}&  F^2 - 4 b  \lambda \sigma^{\nu} \bar{\lambda} \partial^{\mu} F_{\mu \nu}+ 4b^2 \left[ \vphantom{\frac12} \bar{\lambda}^2 \lambda \Box \lambda + \lambda^2 \bar{\lambda} \Box \bar{\lambda} \right.  \\[-10pt]
&& \left.+ \frac 12 \partial_{\mu}(\lambda^2)\partial^{\mu}(\bar{\lambda}^2)  +(\ldbl)(\lbdbl) - \frac 12 \left (( \lambda \slashed{\partial} \bar{\lambda})^2 + (\lbdbl)^2\right) \right] + \text{total derivatives},\nonumber
\eeqn
and cancels the dimension-6 operator coming from \eqref{gauginotransfieldredef1}. This field redefinition acts on other terms in the Lagrangian as follows
\begin{flalign}
&\hspace{1.5cm}F^4 \underset{\eqref{fielddefbosonstrange}}{\longrightarrow}  F^4+2b\partial^{\nu}(\lsl)F_{\nu\mu} F^2+\cdots ~,&\\[5pt]
&\hspace{1.5cm}(F\Ft)^2 \underset{\eqref{fielddefbosonstrange}}{\longrightarrow} (F\Ft)^2+2b\partial^{\nu}(\lsl)\Ft_{\nu\mu} F\Ft+\cdots~, &\\
 \label{transdim6boson}
&\hspace{1.5cm}\lsl \partial^{\nu}F_{\nu \mu}\underset{\eqref{fielddefbosonstrange}}{\longrightarrow}\lsl \partial^{\nu}F_{\nu \mu} - b \partial_{\rho}(\lambda^2) \partial^{\rho}(\bar{\lambda}^2) -  b \partial_{\mu} (\lsl) \partial_{\rho}(\lambda \sigma^{\rho}\bar{\lambda}) + \cdots~, &\\
\label{dim10termsfromdim8vanishing}
 &\hspace{1.5cm}i \lsl \Ft_{\nu \mu} \p_{\rho}F^{\rho \nu}  \underset{\eqref{fielddefbosonstrange}}{\longrightarrow}  i \lsl \Ft_{\nu \mu} \p_{\rho}F^{\rho \nu} + \left(  b i \ldb^2 \ld {\sigma}^{ab}\p_{\rho}\ld \,\partial^{\rho}F_{ab} + {\rm c.c.}\right) +\cdots~, &\\
 \label{eqdim10froml2F2}
&\hspace{1.5cm} i \lambda \sigma^{\mu} \partial_{\rho} \bar{\lambda} F_{\mu\nu} F^{\nu \rho } +{\rm c.c.} \underset{\eqref{fielddefbosonstrange}}{\longrightarrow}i \lambda \sigma^{\mu} \partial_{\rho} \bar{\lambda} F_{\mu\nu} F^{\nu \rho } + ib \ldb^2 \, \ld \sigma^{ab}\p^{\mu} \ld \, \p_{\mu}F_{ab}+ {\rm c.c.}+\cdots~. &
\end{flalign}
After applying the field redefinitions \raisebox{.5pt}{\textcircled{\raisebox{-.9pt} {1}}} and \raisebox{.5pt}{\textcircled{\raisebox{-.9pt} {2}}}, the dimension-6 terms are eliminated completely and  the Lagrangian becomes
\beqn\label{Lagrangianformdim12NL3}
 \mathcal{L}_{\rm SDBI+FI}&\rightarrow& A \left(\vphantom{A^{a'}}1+ \kp i \ldbl +  \kp i \lbdbl \right) + \left(A\kpp+4{Bb^2}\kappa^2 + \frac C2b\kp\right) \partial^{\mu}(\lambda^2)\partial_{\mu}(\bar{\lambda}^2)\nonumber\\[5pt]
 &&+\# \kp(\ld \slashed{\partial} \bar{\ld})(\lbdbl) - \#  \kp( \ld \slashed{\partial} \bar{\ld})^2 - \#  \kp(\lbdbl)^2  + \#  \kp\bar{\ld}^2 \ld \Box \ld + \#  \kp\ld^2 \bar{\ld} \Box \bar{\ld}  \nonumber\\[5pt]
  &&+~ 2A   {a}^2 \kp\,i \lsl \Ft_{~\nu \mu} \p_{\rho}F^{\rho \nu}+2 B \kp \left({F^2}+\kp (F\Ft)^2-\kappa^2 F^4\right) \nonumber\\[5pt] 
&&+ \left(2{B}{\kappa^2}+\frac{A{a}^2}{2}-\frac{Ca}2\right)\kp i\ldbl F^2 -\frac{C{a}}4 \kp i \ldbl F\Ft + {\rm c.c.} \nonumber\\[2pt] 
&&+\left(8{B}{\kappa^2}-\frac{C^2}{8A}\right)\kp  i \lambda \sigma^{\mu} \partial_{\rho} \bar{\lambda}   F_{\mu\nu} F^{\nu \rho } + {\rm c.c.}\nonumber\\[2pt]
&&+~ 8Bb\kappa^4\ld\sigma^{\mu}\partial^{\nu}\ldb~\left(\Ft_{\mu\nu}F^2+F_{\mu\nu}F^2-\Ft_{\mu\nu}F\Ft-F_{\mu\nu}F\Ft\right) + {\rm c.c.}~.\nonumber\\[5pt]
&&-~16{B}{\kappa^6}( i \lambda \sigma^{\mu} \partial_{\rho} \bar{\lambda} - i \partial_{\rho} \lambda \sigma^{\mu} \bar{\lambda})  F_{\mu\nu} F^{\nu \rho }F^2 + 4{B}{a}^2{\kappa^4} i\lsl F^2\partial_{\mu}(F\Ft) +\cdots~. \hspace{1cm}
\eeqn
Note  that the four-fermion/one-gauge-boson dimension-10 terms coming from \eqref{dim10fromdetA}, \eqref{dim10termsfromdim8vanishing} and \eqref{eqdim10froml2F2} cancel each other. They are thus absent in Lagrangian \eqref{Lagrangianformdim12NL3}.


\paragraph{Field redefinition \raisebox{.5pt}{\textcircled{\raisebox{-.9pt} {3}}} } The second line of  \eqref{Lagrangianformdim12NL3} contains goldstino self interactions whose coefficients are not shown explicitly. They can actually be removed completely by applying the field redefinition\,\footnote{Field redefinitions in the form of \eqref{transgoldstino} were used in \cite{Kuzenko:2010ef} to demonstrate in components the on-shell equivalence of different goldstino Lagrangians.}
\be \label{transgoldstino}
\raisebox{.5pt}{\textcircled{\raisebox{-.9pt} {3}}}\quad \ld_{\alpha}\rightarrow  \lambda_{\alpha}+ m \lambda_{\alpha} (\lambda \slashed{\partial}\bar{\lambda})-n \lambda_{\alpha} (\lbdbl)+p\lambda\sigma^{\mu}\bar{\lambda}\partial_{\mu}\lambda_{\alpha}~,\qquad [m]=[n]=[p]=4\,,
\ee
under which kinetic terms transform as
\beqn \label{fieldredefVA}
i\ldbl + i \lbdbl &\underset{\eqref{transgoldstino}}{\longrightarrow}& i\ldbl + i \lbdbl+ 2 i  (m - p)(\ldbl)^2 - 2 i (\bar{m}-\bar{p}) (\lbdbl)^2 + i p \lambda^2 \bar{\lambda} \Box \bar{\lambda}  -i {\bar{p}} \bar{\lambda}^2 {\lambda} \Box {\lambda}\nonumber \\[-5pt]
&& \hspace{-0.2cm}- 2 i (n - \bar{n} - p+ \bar{p}) (\ldbl)(\lbdbl) + \text{total derivatives} + O(\lambda^6).\hspace{1cm} 
\eeqn
We see that there are enough parameters in \eqref{fieldredefVA} to cancel all four-fermion terms except for $\p_{\mu}(\ldb^2)\p^{\mu}(\ld^2)$ which is thus the only physical dimension-8 contribution to the Volkov-Akulov Lagrangian. Under field redefinition  \raisebox{.5pt}{\textcircled{\raisebox{-.9pt} {3}}}, other   terms in the Lagrangian generate dimension-12 terms of the form $\p^2\ld^4F^2$,  or terms with dimension higher than 12.

\paragraph{Field redefinition \raisebox{.5pt}{\textcircled{\raisebox{-.9pt} {4}}} } The first term in the third line of \eqref{Lagrangianformdim12NL3} is proportional to the equation of motion of a free gauge boson  and thus can be eliminated. This is realized by using the field redefinition of  the gauge boson
\beqn\label{fieldredefuhighorder}
\raisebox{.5pt}{\textcircled{\raisebox{-.9pt} {4}}} \quad u_{\mu} \rightarrow u_{\mu} + f i\ld\sigma^{\rho}\ldb \Ft_{\mu\rho}, \qquad f=\frac {Aa^2}{4B}\,,\quad [f]=-4\,,
\eeqn
or equivalently the field redefinition of  the gauge  field strength
\beqn \label{fieldredefFhighorder}
F_{\mu\nu} \rightarrow F_{\mu\nu}+2f\p_{[\mu}\left(i\ld\sigma^{\rho}\ldb \Ft_{\nu]\rho}\right)~.
\eeqn
Under this redefinition the gauge boson kinetic term becomes
\beqn
F^2 &\underset{\eqref{fieldredefFhighorder}}{\longrightarrow}& F^2- 4 f i \ld\sigma^{\rho}\ldb \Ft_{\nu\rho} \p_{\mu}F^{\mu\nu} + \cdots ~,
\eeqn
and thus the second term cancels with the first  term in the third line of \eqref{Lagrangianformdim12NL3}. The  field redefinition \eqref{fieldredefuhighorder} also acts on other terms
\beqn
&&F^4 \underset{\eqref{fieldredefFhighorder}}{\longrightarrow}  F^4+2f i \lsl F^2\partial_{\mu}(F\Ft)+\cdots~,\\[5pt]
&&(F\Ft)^2 \underset{\eqref{fieldredefFhighorder}}{\longrightarrow} (F\Ft)^2+2f i \lsl F^2\partial_{\mu}(F\Ft)-8f i\partial_{\mu}(\ld\sigma^{\nu}\ldb)F^{\mu\rho}F_{\rho \nu} F\Ft + \cdots~.
\eeqn
Hence after applying the field redefinitions \raisebox{.5pt}{\textcircled{\raisebox{-.9pt} {3}}} and \raisebox{.5pt}{\textcircled{\raisebox{-.9pt} {4}}},  the Lagrangian further reduces to
\beqn\label{Lagrangianformdim12NL4}
 \mathcal{L}_{\rm SDBI+FI}&\rightarrow& A \left(\vphantom{A^{a'}}1+\kp i\ldbl +\kp i \lbdbl\right) +  \left(A\kpp+4{Bb^2}\kappa^2 + \frac C2 b \kp\right)   \partial^{\mu}(\lambda^2)\partial_{\mu}(\bar{\lambda}^2)\nonumber\\[2pt]
  &&+ \,2 B \kp \left({F^2}+\kappa^2 (F\Ft)^2-\kappa^2 F^4\right) + \left(8{B}{\kappa^2}-\frac{C^2}{8A}\right)\kp  i \lambda \sigma^{\mu} \partial_{\rho} \bar{\lambda}   F_{\mu\nu} F^{\nu \rho } + {\rm c.c.} \nonumber\\[2pt] 
&&+ \left(2{B}{\kappa^2}+\frac{A{a}^2}{2}-\frac{Ca}2\right)\kp i\ldbl F^2 + {\rm c.c.} -\frac{C{a}}4\kp i\ldbl F\Ft  + {\rm c.c.} \nonumber\\[3pt]
&&-\,16{B}{\kappa^6} i \lambda \sigma^{\mu} \partial_{\rho} \bar{\lambda}  F_{\mu\nu} F^{\nu \rho }F^2 + {\rm c.c.}\nonumber\\[3pt]
&&+\, 8Bb\, \kappa^4\ld\sigma^{\mu}\partial^{\nu}\ldb~\left(\Ft_{\mu\nu}F^2+F_{\mu\nu}F^2-\Ft_{\mu\nu}F\Ft-F_{\mu\nu}F\Ft\right)+ {\rm c.c.}\nonumber\\[3pt]
&& + \left(4{B}{a}^2+8{B}f\right)\kpp i\lsl F^2\partial_{\mu}(F\Ft)-16f{B }{\kappa^4}i\partial_{\nu}(\lsl)F^{\nu\rho}F_{\rho\mu}F\Ft+\cdots~.  \hspace{1cm}
\eeqn

\paragraph{Field redefinition \raisebox{.5pt}{\textcircled{\raisebox{-.9pt} {5}}} }The dimension-10 terms in the fifth line of \eqref{Lagrangianformdim12NL4} arrange in such  a way that they are eliminated through the field redefinition
\beqn\label{fieldrederfdim10}
\raisebox{.5pt}{\textcircled{\raisebox{-.9pt} {5}}} \quad \lambda_{\alpha}\rightarrow\lambda_{\alpha}+(\sigma^{\mu\nu}\lambda)_{\alpha}h\left(F_{\mu\nu}F^2- F_{\mu\nu}F\Ft \right), \qquad h=-i \frac{4B b}{A}\kappa^2 \in i\mathbb{R}, \qquad [h]=-6\,.
\eeqn
Indeed, just like \eqref{fieldredfandb}, one can replace $F_{\mu\nu}$ with $\Ft_{\mu\nu}$ in \eqref{fieldrederfdim10}, due to the identity $\sigma^{\rho\gamma}=\frac i2\epsilon^{\rho\gamma\mu\nu} \sigma_{\mu\nu}$.
The goldstino kinetic terms transform under \eqref{fieldrederfdim10} as
\beqn \label{gauginokineticunderfieldredef10}
&& i\ldbl + i\lbdbl 
\underset{\eqref{fieldrederfdim10}}{\longrightarrow}
  i\ldbl + i\lbdbl-2\left(ih\ld\sigma^{\mu}\partial^{\nu}\ldb-i\bar{h}\partial^{\nu}\ld\sigma^{\mu}\ldb\right)\left(F_{\mu\nu}F^2-\Ft_{\mu\nu}F\Ft\right) \nonumber\\
&&\hspace{4.85cm}-\,2\left(ih\ld\sigma^{\mu}\partial^{\nu}\ldb + i\bar{h}\partial^{\nu}\ld\sigma^{\mu}\ldb\right)\left(\Ft_{\mu\nu}F^2-F_{\mu\nu}F\Ft\right),
\eeqn
and cancel exactly  with the dimension-10 terms of \eqref{Lagrangianformdim12NL4}. Acting \eqref{fieldrederfdim10}  on  other terms in  the Lagrangian,  we only get  dimension-14 or dimension-16 terms. 

Therefore, no dimension-10 operator survives in the physical on-shell Lagrangian.  

\paragraph{Field redefinition \raisebox{.5pt}{\textcircled{\raisebox{-.9pt} {6}}} }  We are still left with the dimension-8 terms of the form $\ldbl F\Ft$ and $\ldbl F^2$ in the third line of \eqref{Lagrangianformdim12NL4}. The first can be eliminated through the field redefinition
\beqn\label{fieldredef6}
\raisebox{.5pt}{\textcircled{\raisebox{-.9pt} {6}}} \quad \ld_{\alpha}&\rightarrow&\ld_{\alpha}+c\ld_{\alpha} F\Ft , \qquad c= \frac{C{a}}{8A} \in \mathbb{R} \,, \qquad [c]=-4\,,
\eeqn
which acts on the fermion kinetic terms as
\beqn\label{transkineticdim12a}
i\ldbl + {\rm c.c.} &\underset{\eqref{fieldredef6}}{\longrightarrow}& i\ldbl -c i \lsl \partial_{\mu}(F\Ft)-ic^2\ld\sigma^{\mu}\partial_{\mu}(\ld F\Ft)F\Ft+{\rm c.c.} \nonumber \\
& &=i\ldbl+2 c i\ldbl  F\Ft +{\rm c.c.} +\cdots~,  \hspace{1cm}
\eeqn
and thus  eliminates the   dimension-8 terms containing $\ldbl  F\Ft$. The field redefinition \eqref{fieldredef6} also acts on other terms as
      \beqn\label{transllFFdim12a}
  i \lambda \sigma^{\mu} \partial_{\rho} \bar{\lambda}  F_{\mu\nu} F^{\nu \rho } + {\rm c.c.} 
  &\! \! \! \! \! \!   \underset{\eqref{fieldredef6}}{\longrightarrow}\! \! \! \! \! \!  
  &  i \lambda \sigma^{\mu} \partial_{\rho} \bar{\lambda}  F_{\mu\nu} F^{\nu \rho }-ic\lsl\partial_{\nu}(F\Ft)F_{\mu\rho}F^{\rho\nu}+ {\rm c.c.}+\cdots \\ [-3pt]
    &  &\! \! \! \! \!  \! \! \! \!\! \!
     = i \lambda \sigma^{\mu} \partial_{\rho} \bar{\lambda}  F_{\mu\nu} F^{\nu \rho } +ic\partial_{\nu}(\lsl)F\Ft F_{\mu\rho}F^{\rho\nu} +\frac{ic}{4}\lsl F^2\partial_{\mu}(F\Ft)+ {\rm c.c.} +\cdots\,,\hspace{-0.3cm}  
     \nonumber
   \eeqn
     \begin{flalign}
      &\hspace{0.2cm} i\ldbl F^2 + {\rm c.c.} \underset{\eqref{fieldredef6}}{\longrightarrow}  i\ldbl F^2 -ic \lsl F^2\partial_{\mu}(F\Ft) +{\rm c.c.}+  \cdots~,& \\
   &\hspace{0.2cm}  i\ldbl F\Ft + {\rm c.c.}\underset{\eqref{fieldredef6}}{\longrightarrow}  i\ldbl F\Ft + {\rm c.c.}+\cdots~.  &
\end{flalign}
     To get to the last line of \eqref{transllFFdim12a} we integrated by part, used Bianchi identities of $F$, as well as its antisymmetry. 

\paragraph{Field redefinition \raisebox{.5pt}{\textcircled{\raisebox{-.9pt} {7}}} }The other dimension-8 operator $\ldbl F^2$  can be eliminated by the following field redefinition
\beqn \label{fieldredef7}
\raisebox{.5pt}{\textcircled{\raisebox{-.9pt} {7}}} \quad \ld_{\alpha}&\rightarrow&\ld_{\alpha}+e\ld_{\alpha} F^2, \qquad e= -\frac{B}{A}{\kappa^2}-\frac{{a}^2}{4}+\frac{Ca}{4A} \in \mathbb{R}\,, \qquad[e]=-4\,.\label{fieldtransfordim12b}
\eeqn
Indeed for $e\in \mathbb{R}$, the fermion kinetic terms transform as
\beqn\label{transkineticdim12b}
i\ldbl+i\lbdbl &\underset{\eqref{fieldtransfordim12b}}{\longrightarrow}&i\ldbl+i\lbdbl  +2e (i\ldbl+i\lbdbl)F^2+\left[ie^2\ld\sigma^{\mu}\partial_{\mu}(\ld F^2)F^2+{\rm c.c.}\right]\nonumber \\
& &= i\ldbl+i\lbdbl+2e (i\ldbl+i\lbdbl)F^2 + \cdots~,
\eeqn
and implements the desired cancellation. The dimension-8 term itself transforms under \eqref{fieldtransfordim12b} as
\beqn
\label{transdim12inLb}
  i \lambda \sigma^{\mu} \partial_{\rho} \bar{\lambda} F_{\mu\nu} F^{\nu \rho } +{\rm c.c.} &\underset{\eqref{fieldtransfordim12b}}{\longrightarrow}&
  i \lambda \sigma^{\mu} \partial_{\rho} \bar{\lambda} F_{\mu\nu} F^{\nu \rho }+ 2e i\lambda \sigma^{\mu} \partial_{\rho} \bar{\lambda} F_{\mu\nu} F^{\nu \rho }F^2 + {\rm c.c.} +\cdots~.
\eeqn
Therefore, under combined field redefinitions \raisebox{.5pt}{\textcircled{\raisebox{-.9pt} {5}}}, \raisebox{.5pt}{\textcircled{\raisebox{-.9pt} {6}}} and \raisebox{.5pt}{\textcircled{\raisebox{-.9pt} {7}}} in the Lagrangian \eqref{Lagrangianformdim12NL4}, we arrive at
\beqn \label{Lagrangianformdim12NL6}
  \mathcal{L}_{\rm SDBI+FI}\rightarrow&&\hspace{-13pt} A + A \kp(i\ldbl + i \lbdbl) +   \left(A\kpp+4Bb^2\kappa^2 + \frac C2b\kp\right)   \partial_{\mu}(\lambda^2)\partial^{\mu}(\bar{\lambda}^2) \nonumber \\
&&\hspace{-18pt} +2 B\kp \left({F^2}+\kappa^2 (F\Ft)^2-\kappa^2F^4\right) + \left(8{B}{\kappa^2}-\frac{C^2}{8A}\right)\kp\left( i \lambda \sigma^{\mu} \partial_{\rho} \bar{\lambda} - i \partial_{\rho} \lambda \sigma^{\mu} \bar{\lambda}\right)  F_{\mu\nu} F^{\nu \rho } \nonumber\\
&&\hspace{-18pt} -\left(16{B}{\kappa^4}+\left(8{B}{\kappa^2}-\frac{C^2}{8A}\right)\left(\frac {2B}A {\kappa^2}+\frac{{a}^2}{2}-\frac{Ca}{2A}\right)\right)\kp\left ( i \lambda \sigma^{\mu} \partial_{\rho} \bar{\lambda} - i \partial_{\rho} \lambda \sigma^{\mu} \bar{\lambda}\right)  F_{\mu\nu} F^{\nu \rho }F^2 \nonumber \\[5pt]
&&\hspace{-18pt}+O(\ld^2F^6)+ O(\ld^4F^2)\,.
\eeqn
\paragraph{Rescaling and final on-shell Lagrangian.} We see that most of the dimension-12 terms cancelled in the Lagrangian \eqref{Lagrangianformdim12NL6}. Kinetic terms can be brought to standard normalizations through the following rescaling \beqn\label{fieldrenormalizationhere}
\lambda \bar{\lambda} &\rightarrow& -\frac {\lambda \bar{\lambda}}{2A\kp g^2} ~, \\
F^2 &\rightarrow& - ~\frac{F^2}{8B\kappa^2g^2}\,~.
\eeqn
Using expressions \eqref{fieldredfandb}, \eqref{fielddefbosonstrange} for $a$, $b$ and \eqref{defABC}  for $A,B$ and $C$ and defining the new constant $\bar{\kappa}$  
\beqn \label{constantsdefinedhere}
\bar{\kappa}^2 \equiv  \frac{\kappa^2}{\sqrt{1+16g^4\kappa^2\xi^2}}~,\eeqn
we can rewrite the Lagrangian \eqref{Lagrangianformdim12NL6} in a much simpler way\,\footnote{Changing the sign  of the first term ``1" in \eqref{Lnonlin} and thus in the definition of $A$ in \eqref{defABC}, following Ref.~\cite{Cribiori:2018dlc}, one gets the same on-shell action up to the order we consider, appart from the cosmological constant term which has no role in global supersymmetry. However, in the limit $\xi =0$ the field redefinitions \eqref{fieldrederfdim10} and \eqref{fieldredef7} become singular because $A$ vanishes.}
\beqn \label{Lagrangianformdim12NL6onshell}
\mathcal{L}_{\rm SDBI+FI}&=& -~ \frac{1}{8 \kappa^2 g^2} \left(1+\frac{\kappa^2}{\bar{\kappa}^2}\right) - \frac 1{2g^2} (i\ldbl + i \lbdbl) - \frac {\bar{\kappa}^2}{g^2} ~ \partial_{\mu}(\lambda^2)\partial^{\mu}(\bar{\lambda}^2) \nonumber \\
&& - ~\frac{F^2}{4g^2} - \frac { \bar{\kappa}^2}{4g^2} \left((F\Ft)^2-F^4 \vphantom{{1'}^2} \right)- \frac { 2  \bar{\kappa}^2}{g^2} \left( i \lambda \sigma^{\mu} \partial_{\rho} \bar{\lambda} - i \partial_{\rho} \lambda \sigma^{\mu} \bar{\lambda}\right)  F_{\mu\nu} F^{\nu \rho } \nonumber \\
&& + ~\frac { 6  \bar{\kappa}^4}{g^2} \left(i \lambda \sigma^{\mu} \partial_{\rho} \bar{\lambda} - i \partial_{\rho} \lambda \sigma^{\mu} \bar{\lambda}\right)  F_{\mu\nu} F^{\nu \rho }F^2+ O\left((\lambda\p\bar\lambda)^2 F^2\right)+O({\rm dim} ~14).
\eeqn
Below are a few comments on the dimension-8 operators present in \eqref{Lagrangianformdim12NL6onshell}. The four-fermion term in the first line corresponds to the expansion of the Volkov-Akulov (VA) action with the redefined decay constant $\bar{\kappa}$. The $F^4$ in the second line corresponds to the expansion of the bosonic DBI action with the same redefined tension. The two-fermion two-boson term in the second line is a consequence of the low energy theorem for the goldstino coupling to matter which 
to leading order is given by $\left(i \lambda \sigma_{\mu} \partial_{\nu} \bar{\lambda} - i \partial_{\nu } \lambda \sigma_{\mu} \bar{\lambda}\right) T^{\mu\nu}$. Here, the stress-energy tensor of the bosonic DBI action is $T^{\mu\nu}=  F^{\mu  \lambda} F^{\nu}_{\,\,\,\,\lambda}  -\frac14 \eta^{\mu\nu}F^2+\cdots$. The trace part  $ \eta^{\mu\nu}F^2$ vanishes on-shell, hence
to leading order we are left with the dimension-8 operator at the end of the second line in \eqref{Lagrangianformdim12NL6onshell}. The dimension-12 term in the third line can also be explained in a similar way. Nevertheless, the relative coefficient between the bosonic DBI action and the fermionic terms, as well as the value of $\bar\kappa$ cannot be obtained from the low energy theorem.

%
%
%
%
%

To summarize, by applying the following series of field redefinitions on  \eqref{Lnonlin}, 
\beqn
\ld_{\alpha}&\underset{ \raisebox{.5pt}{\scriptsize \textcircled{\raisebox{-.8pt} {1}}} \raisebox{.5pt}{\scriptsize\textcircled{\raisebox{-.9pt} {3}}} \raisebox{.5pt}{\scriptsize\textcircled{\raisebox{-.9pt} {5}}} \raisebox{.5pt}{\scriptsize\textcircled{\raisebox{-.9pt} {6}}} \raisebox{.5pt}{\scriptsize\textcircled{\raisebox{-.9pt} {7}}}}{\longrightarrow}& \sqrt{-\frac {1}{2Ag^2}} \left(\vphantom{\frac12}\lambda_{\alpha}+ ia (\sigma^{\mu \nu} \lambda )_{\alpha} F_{\mu \nu}+ m \lambda_{\alpha} (\lambda \slashed{\partial}\bar{\lambda})-n \lambda_{\alpha} (\lbdbl)+p\lambda\sigma^{\mu}\bar{\lambda}\partial_{\mu}\lambda_{\alpha} \right.\nonumber\\
&& \hspace{55pt}+\left.c\ld_{\alpha} F\Ft \label{fieldtransfordim12a}+e\ld_{\alpha} F^2 +h(\sigma^{\mu\nu}\lambda)_{\alpha}\left(F_{\mu\nu}F^2- F_{\mu\nu}F\Ft \right) \vphantom{\frac12}\right),\\[10pt]
u_{\mu} &\underset{ \raisebox{.5pt}{\scriptsize\textcircled{\raisebox{-.8pt} {2}}} \raisebox{.5pt}{\scriptsize\textcircled{\raisebox{-.8pt} {4}}}}{\longrightarrow}& - ~\frac{1}{8\kappa^2g^2B} \left( u_{\mu} +b \lambda \sigma_{\mu} \bar{\lambda}~+ f \ld\sigma^{\rho}\ldb \Ft_{\mu\rho} \vphantom{\frac12}\right),
\eeqn
 we arrive at the low energy on-shell Lagragian~\eqref{Lagrangianformdim12NL6onshell}.

The on-shell Lagrangian~\eqref{Lagrangianformdim12NL6onshell} has the same functional form whenever the FI parameter $\xi$  is zero or not, except for the trivial constant piece. It follows that the FI parameter $\xi$ enters the on-shell Lagrangian only through the renormalization of the coupling constant $\kappa$.  This suggests that  \eqref{Lnonlin} is on-shell equivalent to 
\beqn \label{nonlinLagrangianfinal}
\mathcal{L}^'_{\rm SDBI+FI}&=& -~ \frac{1}{8 \kappa^2 g^2} \left(1-\frac{\kappa^2}{\bar{\kappa}^2}\right)   - \frac{1}{8 g^2 \bar\kappa^2}  \det{\bm A} \left(1+\sqrt{1+4\bar\kappa^2 \mathcal{F}^2+4\bar\kappa^4 (\mathcal{F}\tilde{\mathcal{F}})^2}\right) \nonumber\\
&=&-~ \frac{1}{8 \kappa^2 g^2} \left(1-\frac{\kappa^2}{\bar{\kappa}^2}\right)  -  \frac{1}{8 g^2 \bar\kappa^2} \det{\bm A} \left(1+\sqrt{-\det\left(\eta_{\mu\nu}+2\sqrt2\bar\kappa \mathcal{F}_{\mu\nu}\right)} \right).
\eeqn

It is easy to verify  that by setting  $\lambda=0$ and thus $\det{\bm A}=1$, \eqref{nonlinLagrangianfinal} agrees with the 
bosonic truncation of the   SDBI+FI model~\eqref{Lnonlin}.
In the purely fermionic case $F=0$,  \eqref{nonlinLagrangianfinal} is   reduced to the VA action. This is also consistent with the well-known fact that the VA action provides the low energy description of the supersymmetry breaking.  Together with our explicit computations, the above results provide strong evidence that \eqref{nonlinLagrangianfinal} is 
equivalent to  \eqref{Lnonlin} on-shell.   So the standard FI term plays a trivial role in the SDBI action  by just redefining the coupling constant.

\section{SDBI action with induced FI term from $\gamma$ deformation}\label{sectionexpansion}

In the previous section, we started with the non-linear supersymmetry representation of SDBI+FI model derived in \cite{Cribiori:2018dlc}, considered its low energy expansion and obtained the on-shell physical Lagrangian with both bosons and fermions up to order of dimension 12 (the latter operators involving two gauginos). The non-linear supersymmetry formalism makes the non-linear supersymmetry of SDBI action explicit. However, the linear supersymmetry is obscure and rather invisible. 

In this section, we start with the linear supersymmetry representation of SDBI action or its  generalization SDBI+$\gamma,\theta$ with manifest $\mathcal N=1$ supersymmetry, and then compute the on-shell physical Lagrangian  by means of field redefinitions. The final result of our computations confirms what we obtained in the previous section based on non-linear supersymmetry formalism. 
 
Below, in \cref{SDBIfromconstraint}, we first review briefly the construction of SDBI+$\gamma,\theta$ action and discuss how to recover the SDBI+FI as a particular limit. Then, we expand the action up to operators of dimension 10 (included) in \cref{componentexpansion}  and compute the  on-shell physical Lagrangians of SDBI+$\gamma,\theta$ and SDBI+FI through field redefinitions in  \cref{sdbiwithgamma}.

 \subsection{SDBI +$\gamma,\theta$ action from a non-linear constraint}\label{SDBIfromconstraint}

 The SDBI action or its generalization SDBI+$\gamma,\theta$ can be obtained from the  $\mathcal{N}=2$ vector multiplet $\mathcal{W}$, which consists of a vector multiplet $W$ and a chiral multiplet  $ X$ in $\mathcal N=1$ language. By deforming the   $\mathcal{N}=2$ vector multiplet and imposing the nilpotent constraint  $ \mathcal{W}^2=0 $, the $\mathcal{N}=2$  supersymmetry is partially broken to $\mathcal{N}=1$ and the resulting model leads to the   SDBI   or  SDBI+$\gamma,\theta$ actions. For detailed construction, see \cite{Rocek:1997hi, Antoniadis:2008uk, Antoniadis:2019gbd}.

The constraint  $ \mathcal{W}^2=0 $ can be solved~\cite{Bagger:1996wp}  
   \be\label{XWconstraint}
 X=\kappa W^2 -\kappa^3 \overline D^2 \Big[   \frac{W^2 \overline W^2}{ 1+ \mathcal A +\sqrt{1+2 \mathcal A- \mathcal B^2}}  \Big].
 \ee
Here $W_{\alpha}$ is the usual $\mathcal{N}=1$ gauge field strength superfield contained in $\mathcal{W}$
 \beqn
 \hspace{-0.5cm}W_\alpha = -i \lambda_\alpha +  \theta_\alpha \sfD  -  i (\sigma^{\mu\nu}\theta)_\alpha F_{\mu\nu}    +\taa (\slashed{\p} {\bar \lambda}  )_\alpha.
 \eeqn
It is used to define two superfields
 \be\label{AtBt}
 \mathcal A=\frac{\kappa^2}{2} (D^2 W^2+\overline D^2 \overline W^2 )=\bar {\mathcal A}~ , \qquad 
 \mathcal B=i\frac{\kappa^2}{2} (D^2 W^2-\overline D^2 \overline W^2 )= \bar {\mathcal B}~.
 \ee
 The SDBI action is then given by the $\sf F$ auxiliary field of $X$ which is invariant under both supersymmetries (up to a total derivative), since any power of $ \mathcal{W}$ vanishes by the nilpotent constraint. 
   \be\label{DBI}
\mathcal L_{\rm SDBI}=\frac{1}{8\pi \kappa}\Imag   \Big(\tau \int d^2 \theta  X \Big) = \frac{1}{4 g^2 \kappa}  \left( \int d^2 \theta  X + {\rm c.c.} \right) -i \frac{\theta}{32\pi^2 \kappa} \left( \int d^2 \theta  X - {\rm c.c.} \right)\,,
 \ee
 where 
\be
\tau=\frac{4\pi i}{g^2} +\frac{\theta }{2\pi}~.
\ee
Using the constraint solution \eqref{XWconstraint}, the chiral half-superspace integral of $X$ reads
\beqn
\frac 1 \kappa \int d^2 \theta  X &=& \int d^2 \theta \left(  W^2 -\kappa^2 \overline D^2 \Big[   \frac{W^2 \overline W^2}{ 1+ \mathcal A +\sqrt{1+2 \mathcal A- \mathcal B^2}}  \Big] \right)\nonumber \\
 &=&   \int d^2 \theta  W^2 + 4 \kappa^2 \int d^2 \theta d^2 \bar{\theta}  \frac{W^2 \overline W^2}{ 1+ \mathcal A +\sqrt{1+2 \mathcal A- \mathcal B^2}} 
\nonumber \\ &=&  
 \int d^2 \theta  W^2 + \frac 4 {\kappa^2} \int d^2 \theta d^2 \bar{\theta} ~ \frac{W^2 \overline W^2}{\vphantom{{1}^2} D^2W^2 \overline{D}^2\overline{W}^2} \left( 1+ \mathcal A - \sqrt{1+2 \mathcal A- \mathcal B^2}\right)
 \eeqn
where we used the definition~\eqref{AtBt}.

For our computational convenience, we introduce the  following chiral superfield $\Phi$ and  real superfield $\mathcal{M}$   \beqn \label{superfieldsdefinition}
 { \Huge{\Phi}}\equiv\frac{W^2}{\vphantom{{1}^2} \overline{D}^2\overline{W}^2}, \qquad
  \mathcal{M}\equiv 1+ \mathcal A - \sqrt{1+2 \mathcal A- \mathcal B^2}, \qquad \overline D_{\dot\alpha} \Phi=0,\qquad
  \overline \cM =\cM .
  \eeqn
The SDBI action can then be written as 
\beqn \label{LDBIequationtodevelop}
\mathcal L_{\rm SDBI} &=&
\frac{1}{4g^2} \int d^2 \theta \, W^2 +\frac{1}{4g^2} \int d^2 \bar \theta \,\overline W^2 
+ \frac 2{g^2\kappa^2} \int d^4 \theta ~ \Large{\Phi} \overline{\Huge{\Phi}} \mathcal{M}~. 
\eeqn
 As shown in \cite{Antoniadis:2019gbd}, its pure bosonic part, $i.e.$ with $\lambda=\bar\lambda=0$, after elimination of the $\sfD$ auxiliary field, is written  as
  \beqn\label{DBIbosonic}
{\mathcal L}_{bosonic}&=&\frac{1 }{8g^2 \kappa^2}
 +\frac{i \theta F\tilde F     }{32 \pi ^2 \left(8 \gamma ^2 \kappa ^2+1\right)}
  -\frac {1}{8 g^2  \kappa  \tilde \kappa } {\sqrt{ 1+\frac{\theta^2 g^4\gamma^2   \tilde \kappa^2}{8\pi^4}}}
   \sqrt{-  \det\Big(\eta_{\mu\nu}+2 \sqrt{2}\tilde  \kappa F_{\mu\nu}\Big) } . \hspace{1cm}
 \eeqn
 where 
  \be
 \tilde \kappa^2= \frac{\kappa^2}{1+8 \gamma^2 \kappa^2}~.
 \ee
 To prepare   for the next subsection, we write down the free Maxwell piece of~\eqref{LDBIequationtodevelop} explicitly 
\beqn\label{freeL4}
\mathcal L_4 &=&
 \frac1{4g^2} (\sfD^2+\bar{\sfD}^2)-\frac 1{2g^2}(i\ldbl+i\lbdbl) -\frac {F^2}{4g^2} +\frac{\theta}{32 \pi^2}\left(iF\Ft+i(\bar{\sfD}^2 -\sfD^2)\right)  ~,
\eeqn
where $\sfD=d+i\gamma$ is complex.

The model obtained above with three deformation parameters will be referred to as SDBI+$\gamma,\theta$ model. The SDBI+FI model we discussed in the previous section arises by  setting   the deformation   parameter $\gamma=0$, and adding the standard FI term $\xi \int d^4 \theta V \propto \xi d$ to \eqref{DBI} and \eqref{LDBIequationtodevelop}.
Actually SDBI+FI can be obtained from SDBI+$\gamma,\theta$.  The last term in \eqref{freeL4} contains $i \theta (\bar{\sfD}^2 -\sfD^2)\sim \gamma\theta d$ which is the standard FI term $\xi d$ with
 \beqn
 \xi \equiv-\frac{\theta\gamma}{8\sqrt{2}\pi^2}~.
 \eeqn
 Moreover, in the limit $\gamma\to 0$, the non-linear third term in \eqref{LDBIequationtodevelop} reduces to the  one in the standard SDBI.  Hence we conclude that SDBI+FI Lagrangian can be obtained  from the SDBI+$\gamma,\theta$ one by taking the double scaling limit:
 \beqn \label{limitcases}
\mathcal{L}_{{\rm SDBI}+\gamma,\theta} \longrightarrow \mathcal{L}_{\rm SDBI+FI}+\text{total derivative}, \qquad {\rm when}  \qquad \left\{
    \begin{array}{l}
       \gamma \rightarrow 0 \\
        \gamma\theta =-{8\sqrt{2}\pi^2}\xi \quad \text{ fixed} .
    \end{array} 
\right.\eeqn
Of course, this limit is ill-defined at the non-perturbative level, since $\theta$ goes to infinity.
 
\subsection{Component expansion}\label{componentexpansion}

We would like to find the physical on-shell action of \eqref{LDBIequationtodevelop}  including both bosonic and fermionic contributions, by performing a low energy perturbative expansion in mass dimension. 
The non-linear interacting piece in  \eqref{LDBIequationtodevelop} is 
\beqn \label{expansionPhiPhibM}
\int d^4 \theta \, \Large{\Phi} \overline{\Huge{\Phi}}\M &=&\left.\Large{\Phi}\overline{\Huge{\Phi}}\right|_0  \left.\M\right|_{\taa\taab} + \left. \Large{\Phi}\overline{\Huge{\Phi}}\right|_\ta  \left.\M\right|_{\ta\taab} + \cdots + \left. \Large{\Phi}\overline{\Huge{\Phi}}\right|_{\taa\taab}  \left.\M\right|_{0} .  
\eeqn
 The relevant superfield expansions are shown in \cref{superfieldexpansion}.
In the following equations, we expand explicitly the various contributions of the superfield multiplication shown in \eqref{expansionPhiPhibM} and keep terms up to  dimension 10\,\footnote{As we will see below, 
the $\sfD$ auxiliary field can be expanded   as $\sfD=\sfD_0+\sfD_4+ \cdots$, where $\sfD_0$ is  constant,  $\sfD_4$ has dimension 4, etc.  Hence the various terms in $\p_{\mu}\sfD$ have dimensions at least 5. }.
Explicitly, the important terms read (we use $\sim$ symbol to indicate that equalities hold up to dimension-10 terms included or total derivatives)
\begin{flalign} \label{developeachterm1}
&\left.\Large{\Phi}\overline{\Huge{\Phi}}\right|_0 \left.\M\right|_{\taa\taab}\sim0~,&  \\ \label{developeachterm2}
& \left.\Large{\Phi}\overline{\Huge{\Phi}}\right|_\tab\left.\M\right|_{\tab\taa}\sim \kappa^2 \ld^2\ldb\Box\ldb\frac 1{8\sfD^2}\left(1-\frac{1+8i\kappa^2d\gamma}{\sqrt{Z}}\right)-\frac{ i\kappa^2 \Db}{4 \sfD^2 \Db^2} \ld^2 \ldb \sigma^{ab} \p_{\mu} \ldb~ \p^{\mu} F_{ab}, &  \\\label{developeachterm3}
& \left.\Large{\Phi}\overline{\Huge{\Phi}}\right|_\taab\left.\M\right|_{\taa}\sim \frac{-\kappa^2 \ld^2\Box(\ldb^2)}{8\sfD^2}\left(1-\frac{1+8i\kappa^2d\gamma}{\sqrt{Z}}\right)+\# \kappa^4(\ldbl)^2
,& \\\label{developeachterm5}
&\left.\Large{\Phi}\overline{\Huge{\Phi}}\right|_{\ta\tab}\left.\M\right|_{\ta\tab}\sim \kappa^2\frac{i{\psi}\sigma^{\mu}\bar{\psi}}{8\sfD^2\Db^2}\left[\p_{\mu}\Big(\cE_-+\sfD^2-\Db^2\Big)\left(1-\frac{1}{\sqrt{Z}}\right)+\p_{\mu}\Big(\cE_+ +\sfD^2+\Db^2\Big)\frac{8i\kappa^2d\gamma}{\sqrt{Z}}\right]&\nonumber\\[5pt]
&\hspace{2.4cm}+\# \kappa^4(\ldbl)(\lbdbl)
, & \\ 
\label{developeachterm6} &\left.\Large{\Phi}\overline{\Huge{\Phi}}\right|_{\taab\ta} \left.\M\right|_{\ta}\sim \kappa^2\frac{i\psi\slashed{\p}\bar{\psi}}{4\sfD^2}\left(1+\frac{{F^+}^2}{\sfD^2}\right)\left[\vphantom{\frac{{F^+}^2}{\sfD^2}} 1-\frac{1+8i\kappa^2d\gamma}{\sqrt{Z}}-2\kappa^2{\cE_-}\left(\frac{1}{\sqrt{Z}}+\frac{8i\kappa^2d\gamma-64\kappa^4d^2\gamma^2}{\G\sG}\right)\right.&\nonumber\\
&\hspace{2.2cm} \left.-2\kappa^2\cE_+\frac{1+8i\kappa^2d\gamma}{\G\sG} \vphantom{\frac{{F^+}^2}{\sfD^2}} \right]-\frac{\kappa^2}{8\sfD^2}\ld^2\ldb\Box\ldb\left(1-\frac{1+8i\kappa^2d\gamma}{\sqrt{Z}}\right) 
+\frac{i\kappa^2\Db}{4\sfD^2\Db^2} \ld^2\ldb \bar{\sigma}^{ab} \p_{\mu} \ldb \p^{\mu}F_{ab} \nonumber&\\[5pt]
&\hspace{2.2cm}+\# \kappa^4(\ldbl)(\lbdbl)+\# \kappa^4(\ldbl)^2, &\\
\label{developeachterm7}
\displaystyle & \left.\Large{\Phi}\overline{\Huge{\Phi}}\right|_{\taa\taab} \left.\M\right|_0 \sim \left(1-4\kappa^2(d^2-\gamma^2)-\sG\right)\left[\frac1{32}+\frac {i\psi\slashed{\p}\bar{\psi}}{16\sfD^2\Db^2}\left(1+\frac{2{F^+}^2}{\sfD^2}\right)+\frac {i\psi\sigma^{\mu}\bar{\psi}}{8\sfD^2\Db^2}\frac{\p_{\mu}\sfD}{\sfD} \right.\nonumber&\\
&\hspace{2.4cm}\left.+~\frac 1{ 32\sfD^2\Db^2} \p_{\mu}(\ld^2)\p^{\mu}(\ldb^2)\vphantom{\frac{\Psi^{2^2}}{\Psi^2}} \right]-\kappa^2 \frac {i \psi\slashed{\p}\bar{\psi}}{8\sfD^2\Db^2}\left(\cE_+\left(1-\frac1{\sG}\right)+\cE_-\frac{8i\kappa^2d\gamma}{\sG}\right)& \nonumber\\
&\hspace{2.4cm}-~\frac{\kappa^4}{16}\left(\cE_-^2\left(\frac1{\sG}+\frac{64\kappa^4d^2\gamma^2}{\G\sG}\right)-\frac{\cE_+^2}{\G\sG}+2\cE_+\cE_-\frac{8i\kappa^2d\gamma}{\G\sG}\right),&\nonumber\\[5pt]
&\hspace{2.4cm}+\# \kappa^4(\ldbl)(\lbdbl)+\# \kappa^4(\ldbl)^2+ {\rm c.c.} ~.& 
\end{flalign}
The (anti-)self-dual tensors $F^{\pm}$ are defined in \eqref{dualtensors} while $\psi$,  $\cE_{\pm}$, and $Z$ are introduced in \eqref{defofPsi}, \eqref{defEpm} and \eqref{labelZ}. In particular, $Z$ is given by
\be \label{Gammafun}
Z =(1+8\kappa^2\gamma^2)(1-8\kappa^2d^2)=1-8\kappa^2\left(d^2-\gamma^2\right)-(8\kappa^2d\gamma)^2.
\ee
In \eqref{developeachterm7} the final c.c. symbol refers to complex conjugation of the whole right-hand side, even if some terms are real by themselves. We show \eqref{developeachterm7} in this form to stress the fact that this term is real. In equations \eqref{developeachterm1} to \eqref{developeachterm7}, we put \# in front of four-gaugino terms to indicate that the corresponding coefficients can be calculated but their specific values are not important. As we explain later, these terms can be eliminated in the end by a field redefinition.

Collecting all the above terms, the SDBI Lagrangian \eqref{LDBIequationtodevelop} can be expanded up to dimension 10 as\,\footnote{Note that the dimension-8 and dimension-10 contributions of the form $\ld^2\ldb\Box\ldb$, $\ld^2\ldb\bar{\sigma}^{ab}\p^{\mu}\ldb\p_{\mu}F_{ab}$ present in \eqref{developeachterm2} and \eqref{developeachterm6} cancel each other in the Lagrangian \eqref{SDBIdeveloppedFinal} (and so do their complex conjugates).   
 }
\begingroup\makeatletter\def\f@size{10}\check@mathfonts
\beqn \label{SDBIdeveloppedFinal}
\mathcal{L}_{{\rm SDBI}+\gamma,\theta}&=&\frac{1}{8\kappa^2g^2}\left(1-\sG\right) + \frac{\theta}{32 \pi^2 }(iF\Ft+4\gamma d)-\frac {F^2}{4 g^2 \sG }+ iF\Ft\frac{2\kappa^2 d \gamma}{g^2\sG} +\frac{\kappa^2F^4}{4g^2\G\sG}  \nonumber \\
&&-i \kappa^2F\Ft F^2 \frac{4 \kappa^2d\gamma}{g^2\G\sG}  -\frac{\kappa^2(F\Ft)^2}{4g^2}\left(\frac 1{\sG}+\frac{64\kappa^4 d^2\gamma^2}{\G\sG}\right) -\left(i\ldbl + {\rm c.c.}\right)\frac{1}{2g^2\sG}\nonumber \\
&&+\frac{i\psi\slashed{\p}\bar{\psi}}{2g^2\sfD^2}\left(1-\frac{1+8i\kappa^2 d\gamma}{\sG}+\frac{1}{4\kappa^2\Db^2}\left(1-4\kappa^2(d^2-\gamma^2)-\sG\right)\right)+ {\rm c.c.} \nonumber \\[3pt]
&&+ \# (\ld \slashed{\partial} \bar{\ld})(\lbdbl) - \# ( \ld \slashed{\partial} \bar{\ld})^2 - \# (\lbdbl)^2\nonumber\\[3pt]
&&+\frac{\p_{\mu}(\ld^2)\p^{\mu}(\ldb^2)}{2g^2}\left[\frac1{2\sfD^2}\left(1-\frac{1+8i\kappa^2d\gamma}{\sG}\right)+ {\rm c.c.} + \frac{1}{4\kappa^2\sfD^2\Db^2}\left(1-4\kappa^2(d^2-\gamma^2)-\sG\right)\right]\nonumber \\
&&+\frac{i\psi\slashed{\p}\bar{\psi}}{2g^2\sfD^2}\left[\left(1-\frac{1+8i\kappa^2d\gamma}{\sG}\right)\frac{{F^+}^2}{\sfD^2}+2\kappa^2F\Ft\left(\frac 1{\sG}+\frac{64\kappa^4d^2\gamma^2-8i\kappa^2d\gamma}{\G\sG}\right)+2\kappa^2F^2\frac{1+8i\kappa^2d\gamma}{\G\sG}\right.\hspace{-1cm}\nonumber\\
&&\left.+\frac 1{2\Db^2}\left(F\Ft+F^2\right)\left(1-\frac{1-8i\kappa^2d\gamma}{\sG}+\frac1{2\kappa^2\sfD^2}\left(1-4\kappa^2(d^2-\gamma^2)-\sG\right)\right)\right]+{\rm c.c.}\nonumber\\
&&-\kappa^2\left(i\ldbl F\Ft\left(\frac 1{\sG}+\frac{64\kappa^4d^2\gamma^2}{\G\sG}\right)-i\ldbl \frac{F^2}{\G\sG}+i\ldbl (F\Ft+F^2)\frac{8i\kappa^2d\gamma}{\G\sG}\right)+{\rm c.c.}\nonumber\\
&&+\kappa^2\frac{i\psi\sigma^{\mu}\bar{\psi}}{4g^2\Db^2\sfD^2}\left(\frac{1}{\sfD}-\frac{1}{\Db}\right)\p_{\mu}d \left(1-4\kappa^2(d^2-\gamma^2)-\sG\right)+{\rm c.c.} -\frac{\psi\sigma^{\mu}\bar{\psi}}{g^2\Db^2\sfD^2}  \gamma \p_{\mu}d\left(1-\frac {1-8 \kappa^2 d^2}{\sG}\right)\hspace{-0.2cm} \nonumber\\[5pt]
 &&+ \text{total derivatives}+O(\text{dim }12)~,
\eeqn
\endgroup
where each c.c. symbol indicates now complex conjugation of the first preceding term. Again,  \#  coefficients  in front of the four-fermion dimension-8 terms are not important since these terms can  be removed through field redefinitions, as explained later.

\subsection{Physical action with $\gamma$ deformation and $\theta$ angle} \label{sdbiwithgamma}
In this subsection we compute the physical (on-shell) SDBI+${\gamma,\theta}$ Lagrangian out of the off-shell one in \eqref{SDBIdeveloppedFinal}.
It is clear from \eqref{psidpsi} that when $\langle\sfD\rangle\neq0$ the SDBI Lagrangian \eqref{SDBIdeveloppedFinal} contains dimension-6 terms. As explained in \cref{subsecphysicalNL}, these terms are unphysical and can be eliminated by means of a field redefinition. This field redefinition generates also extra higher dimensional operators when acting on other terms in the Lagrangian. 

Below, we proceed as follows. We first solve the equation of motion of $\sfD$ and use it to obtain the $\sfD$-solved Lagrangian. Since the Lagrangian is too complicated, we only show explicitly the dimension-6 terms and the gaugino kinetic terms. Then we demonstrate how to eliminate the  dimension-6 terms through   field redefinition and write down the  $\sfD$-solved Lagrangian without dimension-6 terms.  After discussing the elimination of some other unphysical terms,   we finally  obtain the physical  on-shell action  of   SDBI+${\gamma,\theta}$ Lagrangian up to dimension 10 (included). 
 
\paragraph{Solving the $\sfD$ auxiliary field.} 
In the presence of $\gamma$ deformation the $\sfD$ auxiliary field splits into its dynamic part $d$ and deformed part $\gamma$ \cite{Antoniadis:2019gbd}. As already mentioned, it is no more real and we get
 \beqn
 \sfD=d+i\gamma, \qquad \Db=d-i\gamma.
 \eeqn
We expand $\sfD$ in terms of increasing dimensions and solve for the first two terms through Euler-Lagrange equations applied in  \eqref{SDBIdeveloppedFinal}. The solution reads
\beqn \label{Dsolutiongammatheta}
&&\sfD\equiv \sfD_0+\sfD_4+\cdots ~,\qquad \sfD_0=d_0+i\gamma, 
\nonumber \\[5pt]
&&d_0=-\frac{\gamma g^2\theta}{2\sqrt{2}\sqrt{8\pi^4+\gamma^2\kappa^2(g^4\theta^2+64\pi^4)}}, \nonumber\\[5pt]
&& \sfD_4=-\frac{2i\kappa^2\gamma F\Ft}{1+8\kappa^2\gamma^2}+d_0\frac{2\kappa^2 F^2}{1+8\kappa^2\gamma^2}+8\kappa^2(d_0-i \gamma)\frac{1+8\kappa^2\gamma^2+2\sqrt{Z_0}}{(1+8\kappa^2\gamma^2+\sqrt{Z_0})^2} (i\ldbl+i\lbdbl),  \nonumber\\[5pt]
 &&Z_0 =(1-8\kappa^2d_0^2)(1+8\kappa^2\gamma^2) =\frac{8\pi^4(1+8 \kappa^2 \gamma^2)^2}{{ \gamma^2\kappa^2 g^4\theta^2+8\pi^4(1+8 \kappa^2 \gamma^2)}} , 
\eeqn
where $Z_0$ is the lowest term in the expansion of $Z$ defined in \eqref{Gammafun}.

We then plug  the above solutions for $\sfD_0$ and $\sfD_4$ back into  \eqref{SDBIdeveloppedFinal}.
Especially, the  $\p_{\mu}d$ factor   in \eqref{SDBIdeveloppedFinal}  can be replaced with $\p_{\mu}\sfD_4$ at dimension-10 order. One can further integrate by parts to transfer the derivative in  $\p_{\mu}\sfD_4$  to other factors. The resulting terms with bare $\sfD_4$ can be combined with other terms in the Lagrangian.
 We do not show the whole $\sfD$-solved Lagrangian but rather present it in schematic form
\beqn\label{DsolvedL}
\mathcal{L}_{\sfD-{\rm solved}}&=&\mathcal{L}_{bosonic} +\mathcal{L}_4\left(\ld\slashed{\p}\ldb,\lbdbl\right)+\mathcal{L}_{6}\left(\ld\sigma^{\mu}\p^{\nu}F_{\mu\nu},\ld\sigma^{\mu}\p^{\nu}\Ft_{\mu\nu}\right)\nonumber\\
&&+\mathcal{L}_{8}\left(\ldbl F^2, \dlbl F^2,\ldbl F\Ft,\dlbl F\Ft, \ld\sigma^{\mu}\p^{\nu} F_{\mu\rho}F^{\rho}_{~\nu}\right)\nonumber\\
&&+\mathcal{L}_{10}\left( \ld\sigma^{\mu}\p^{\nu}F_{\mu\nu}F^2, \ld\sigma^{\mu}\p^{\nu}F_{\mu\nu}F\Ft, \ld\sigma^{\mu}\p^{\nu}\Ft_{\mu\nu}F^2, \ld\sigma^{\mu}\p^{\nu}\Ft_{\mu\nu}F\Ft\right) + \cdots~.\qquad
\eeqn
In the above schematic Lagrangian we indicated the dimension of each term by a subscript, and showed each field dependence (in linearly independent operators up to total derivatives).

The dimension-4 term  $\mathcal{L}_4$   contains the gaugino kinetic terms  
 \beqn\label{L4}
 \mathcal{L}_{4}&=&-\frac{i\ldbl}{2g^2\sGo} + \frac{i\ldbl\sfD_0\Db_0}{2g^2\sfD_0^2}\left(1-\frac{1+8i\kappa^2d_0\gamma}{\sG}+\frac{1}{4\kappa^2\Db_0^2}\left(1-4\kappa^2(d_0^2-\gamma^2)-\sGo\right)\right)+{\rm c.c.} \nonumber \\
&=&-\frac{1+8\kappa^2\gamma^2}{g^2\sGo(1+8\kappa^2+\sGo)}i\ldbl+{\rm c.c.}+ \text{total derivatives}~.
\eeqn
 while  $\mathcal{L}_{6}$ can be obtained by inserting the expression \eqref{psidpsi} into the third line of \eqref{SDBIdeveloppedFinal}, replacing $\sfD \to \sfD_0$ and keeping only the dimension-6 operators
 \beqn \label{L6}
\mathcal{L}_{6}&=&\frac{i\left.\psi\slashed{\p}\bar{\psi}\right|_{6}}{2g^2\Db_0^2\sfD_0^2}\left(\Db_0^2\left(1-\frac{1+8i\kappa^2 d\gamma}{\sGo}\right)+\frac{1}{4\kappa^2}\left(1-4\kappa^2(d_0^2-\gamma^2)-\sGo\right)\right)+ {\rm c.c.} 
\nonumber \\
&=&\frac{-4\kappa^2(1+8\kappa^2\gamma^2)}{g^2\sGo(1+8\kappa^2\gamma^2+\sGo)^2}\left(\vphantom{\frac{2}{3}}2\Db_0\ld\sigma^{\mu}\p_{\nu}\ldb F^{+{\nu}}_{~~~\mu}-2\sfD_0\ld\sigma^{\mu}\p_{\nu}\ldb F^{-{\nu}}_{~~~\mu} -\sfD_0\lsl\p^{\nu}F_{\nu\mu}\right) \nonumber  \\[5pt]
&&+ {\rm c.c.}+\cdots~.
\eeqn

\paragraph {Eliminating dimension-6 terms.} 
The dimension-6 part  $\mathcal{L}_{6}$ shown in \eqref{L6} can be completely eliminated through the field redefinition
\beqn \label{fieldredefhere}
\lambda \rightarrow \ld + a \sigma^{\mu\nu}\ld F_{\mu \nu} ~, \quad a=-i\frac{4\kappa^2}{1+8\kappa^2\gamma^2+\sGo}\Db_0.
\eeqn
Under \eqref{fieldredefhere} the gaugino kinetic term transforms as
\begin{align}
 i\ld \slashed{\partial}\bar{\ld} + {\rm c.c.} &\underset{\eqref{fieldredefhere}}{\longrightarrow} i  \ldbl -2ia \lambda \sigma^{\nu}\partial^{\mu}\bar{\lambda}F^+_{~\mu\nu}-2i \bar{a} \lambda \sigma^{\nu}\partial^{\mu}\bar{\lambda}F^-_{~\mu\nu}  -i\bar{a} \lsl \partial^{\nu}F_{\nu\mu}  \label{transfgauginokinetic} \nonumber \\[-10pt]
 &\hspace{1cm} + ~2 i a\bar{a} \lambda \sigma^{\mu} \partial_{\nu} \bar{\lambda} F_{\mu}^{~\rho}F_{\rho}^{~\nu}+\frac{{a}\bar{a}}2 i\ldbl F^2 + {\rm c.c.}+\cdots~,
\end{align}
and the last three terms in the first line indeed cancel the dimension-6 operator $\mathcal{L}_6$ of \eqref{L6}. The field redefinition \eqref{fieldredefhere} also acts on operators present in $\mathcal{L}_6$ and $\mathcal{L}_8$ as follows \label{actionredef1}
\begin{align}
& i \lambda \sigma^{\mu} \partial_{\nu} \bar{\lambda} F_{\mu\rho} F^{\rho \nu } \underset{\eqref{fieldredefhere}}{\longrightarrow} i \lambda \sigma^{\mu} \partial_{\nu} \bar{\lambda} F_{\mu\rho} F^{\rho \nu}
 +\left( i\frac a4 \lambda \sigma^{\mu} \partial^{\nu} \bar{\lambda} (F_{\nu\mu}-\Ft_{~\nu\mu}) F\Ft+ {\rm c.c.} \right) \label{actionredef1}\nonumber\\[-8pt]
&\hspace{3.5cm} -i\frac{\bar{a}}4 \p_{\nu}(\ld\sigma^{\mu}\ldb)F^{*\nu}_{~~\mu}F^2+i\left(\left(\frac{\bar{a}}4+\frac a2\right)\lambda \sigma^{\mu} \partial^{\nu} \bar{\lambda}-\frac{\bar{a}}4 \partial^{\nu}\lambda \sigma^{\mu} \bar{\lambda} \right)  F_{\mu\nu} F^2+ \cdots~, \\[7pt]
&i\ldbl F^2  \underset{\eqref{fieldredefhere}}{\longrightarrow} i\ldbl F^2  - i(a+\bar{a}) \lambda \sigma^{\mu} \partial^{\nu} \bar{\lambda} F_{\nu\mu} F^2 - i(a-\bar{a}) \lambda \sigma^{\mu} \partial^{\nu} \bar{\lambda} \Ft_{\nu\mu} F^2  +\cdots  ~,\label{actionredef2}\\[5pt]
&i\ldbl F\Ft  \underset{\eqref{fieldredefhere}}{\longrightarrow} i\ldbl F\Ft  - i(a+\bar{a}) \lambda \sigma^{\mu} \partial^{\nu} \bar{\lambda} F_{\nu\mu} F\Ft- i(a-\bar{a}) \lambda \sigma^{\mu} \partial^{\nu} \bar{\lambda} \Ft_{\nu\mu} F\Ft  +\cdots ~, \label{actionredef3}\\[5pt]
&\ld\sigma^{\mu}\partial^{\nu}\ldb \Ft_{\nu\mu}\underset{\eqref{fieldredefhere}}{\longrightarrow} \ld\sigma^{\mu}\partial^{\nu}\ldb \Ft_{~\nu\mu}+\frac a4\ldbl F\Ft + \frac a2 \ldbl F^2 +\left( a\lambda \sigma^{\mu} \partial_{\nu} \bar{\lambda} F_{\mu\rho} F^{\rho \nu } +{\rm c.c.}\right)\nonumber\\[-8pt]
& \hspace{2.7cm} +\left(\frac {a\bar{a}}4 \lambda \sigma^{\mu} \partial^{\nu} \bar{\lambda} (\Ft_{\nu\mu} F^2- F_{\nu\mu} F\Ft+\Ft_{\nu\mu} F\Ft- F_{\nu\mu} F^2) + {\rm c.c.}\right)+ \cdots~, \label{actionredef4} \\[5pt]
&\ld\sigma^{\mu}\partial^{\nu}\ldb F_{\nu\mu}\underset{\eqref{fieldredefhere}}{\longrightarrow} \ld\sigma^{\mu}\partial^{\nu}\ldb F_{\nu\mu}+\frac{a+\bar{a}}4\ldbl F\Ft - \left(a\lambda \sigma^{\mu} \partial_{\nu} \bar{\lambda} F_{\mu\rho} F^{\rho \nu } -{\rm c.c.}\right) \nonumber\\[-8pt]
& \hspace{2.6cm} +\left(\frac {a\bar{a}}4 \lambda \sigma^{\mu} \partial^{\nu} \bar{\lambda} (\Ft_{\nu\mu} F^2- F_{\nu\mu} F\Ft+\Ft_{\nu\mu} F\Ft- F_{\nu\mu} F^2) - {\rm c.c.}\right)+ \cdots~, \label{actionredef5} 
\end{align}
where various identities in appendix~\cref{usefulids} are used and ``$\cdots$" indicate either  total derivatives, terms with dimension higher than 10, or dimension-10 terms proportional to the free equations of motion\,\footnote{As we explained at the beginning of \cref{subsecphysicalNL}, these terms can be eliminated by means of field redefinition without introducing extra terms at this order. 
}.

\paragraph{Eliminating terms containing four fermions.}  We also remark that under the field redefinition \eqref{fieldredefhere} the four-gaugino dimension-8 terms indicated with \# coefficients in \eqref{psidpsi} transform as
\begin{flalign}\label{generated10a}
&\hspace{2cm} (\ldbl)^2 \underset{\eqref{fieldredefhere}}{\longrightarrow}(\ldbl)^2 + \# \ldbl `` \ld \p \ldb F"  + \cdots~, &\\
&\hspace{2cm} (\lbdbl)^2\underset{\eqref{fieldredefhere}}{\longrightarrow}  (\lbdbl)^2+\# \lbdbl ``\overline{ \ld \p \ldb F}" + \cdots~,& \label{generated10b} \\
&\hspace{2cm} (\ldbl)(\lbdbl) \underset{\eqref{fieldredefhere}}{\longrightarrow}  (\ldbl)(\lbdbl)+ \# \ldbl `` \ld \p \ldb F "  + \# \lbdbl `` \overline{\ld \p \ldb F}" + \cdots ~,& \label{generated10c}
\end{flalign}
where $`` \ld \p \ldb F"$ schematically denotes a sum of various contractions containing one field strength, two fermions and one derivative.  Their specific form is not important but the crucial point is that  all these terms are proportional to the equation of a free fermion and thus can be eliminated. 

As we did in section~\ref{sec2NL},  the pure four-fermion terms of dimension 8 in \eqref{generated10a}, \eqref{generated10b},   \eqref{generated10c} can be eliminated through the field redefinition \eqref{transgoldstino}
\begin{flalign} \label{gauginofieldredefmassdim8}
&  \ld_{\alpha}\rightarrow  \lambda_{\alpha}+ m \lambda_{\alpha} (\lambda \slashed{\partial}\bar{\lambda})+n \lambda_{\alpha} (\partial_{\mu}\lambda \sigma^{\mu}\bar{\lambda})+p\lambda\sigma^{\mu}\bar{\lambda}\partial_{\mu}\lambda_{\alpha}~,
\end{flalign}
with constant coefficients $m,n,p$, under which the free fermion kinetic terms transform as \eqref{fieldredefVA}.  It is easy to check that the above four-fermion terms can be eliminated completely with appropriate constants $m,n,p$. Since the dimension-6 terms are eliminated through \eqref{fieldredefhere}, acting \eqref{gauginofieldredefmassdim8} on the remaining terms in the Lagrangian can only generate terms with dimension strictly higher than 10 which we do not consider.

The field redefinition \eqref{fieldredefhere} also generates four fermion and one gauge boson mixed terms with dimension 10 in \eqref{generated10a}, \eqref{generated10b},   \eqref{generated10c}. Since they are proportional to the  equation of free fermion, they can be eliminated through the following schematic field redefinition 
\be\label{gauginofieldredefmassdim10}
 \ld_{\alpha}\rightarrow  \lambda_{\alpha}+ j \lambda_{\alpha} `` \ld \p \ldb F"~. 
\ee
Acting on the free fermion kinetic term, we have
\beqn 
\ldbl\underset{\eqref{gauginofieldredefmassdim10}}{\longrightarrow}\ldbl + j  \ldbl `` \ld \p \ldb F" + \bar{j} \lbdbl``  \overline{\ld \p \ldb F}" +  \cdots~,
\eeqn
which indeed allows us to remove the dimension-10 terms in \eqref{generated10a}, \eqref{generated10b},   \eqref{generated10c} with appropriate tensor structures and coefficients. In particular, no other dimension-10 terms would be generated due to this field redefinition \eqref{gauginofieldredefmassdim10}. 

To conclude, the four fermion terms in \eqref{SDBIdeveloppedFinal} with unspecified coefficients can be completely eliminated at this order without introducing extra terms. The only leftover four fermion operator of dimension 8 is the one written in the fifth line that corresponds to the expansion of the VA action.

 \paragraph{$\sfD$-solved Lagrangian after eliminating dimension-6 operators and non Voklkov-Akulov 4-fermion terms. }
 The field redefinition \eqref{fieldredefhere}  used to eliminate $\mathcal L_6$ also acts on other terms as we see from \eqref{transfgauginokinetic} to \eqref{actionredef5}. Collecting all these terms, we arrive at

\beqn \label{SDBIgammathetawithsolvedD}
\mathcal{L}_{{\rm SDBI}+\gamma,\theta}&=&\frac{1}{8 g^2 \kappa ^2}\left(1-\frac{1+8 \gamma ^2 \kappa ^2}{\sGo}\right) + \frac{i\theta F\Ft}{32 \pi^2 (1+8\kappa^2\gamma^2)}-\frac {F^2}{4 g^2 \sGo } +\frac{F^4-\left( F\Ft\right)^2}{4g^2(1+8\kappa^2\gamma^2)\sGo} \nonumber \\[5pt]
&&- \left(i\ldbl + {\rm c.c.}\right)\frac{1+8\kappa^2\gamma^2}{g^2\sGo(1+8\gamma^2+\sGo)}- \p_{\mu}(\ld^2)\p^{\mu}(\ldb^2)\frac{4\kappa^2 (1+8\kappa^2\gamma^2)}{g^2\sGo(1+8\kappa^2\gamma^2+\sGo)^2} \nonumber\\[5pt]
&&-(i \lambda \sigma^{\rho}\partial_{\nu} \bar{\lambda} -i \partial_{\nu} \lambda \sigma^{\rho} \bar{\lambda})F^{\nu}_{~\mu} F^{\mu}_{~\rho} \frac{4\kappa^2}{ g^2\sGo \left(1+8\kappa^2\gamma^2+\sGo \right)} \nonumber \\[5pt]
&&+~i\ldbl F^2\frac{\kappa^2}{g^2\sGo\left(1+8\kappa^2\gamma^2+\sGo\right)}-i\ldbl F\Ft\frac{2\kappa^2}{g^2\sGo\left(1+8\kappa^2\gamma^2+\sGo\right)}+{\rm c.c.}\nonumber\\[5pt]
&&-~  \ldbl F^2\frac{16 \kappa^4 d_0\gamma(1+8\kappa^2\gamma^2)}{g^2\sGo\left(1+8\kappa^2\gamma^2+\sGo\right)^2}-\ldbl F\Ft\frac{16 d_0\gamma \kappa^4 (1+8\kappa^2 \gamma^2)^2}{g^2\sGo\left(1+8\kappa^2\gamma^2+\sGo\right)^3}+{\rm c.c.}\nonumber\\[5pt]
&&+\frac{8\kappa^4 d_0^2 \quad \ld\sigma^{\mu}\partial^{\nu}\ldb}{g^2\sfD_0\sqrt{Z_0}(1+8\kappa^2\gamma^2+\sqrt{Z_0})^2} \left(\Ft_{~\mu\nu}F\Ft+F_{\mu\nu}F\Ft-\Ft_{~\mu\nu}F^2-F_{\mu\nu}F^2\right) + {\rm c.c.}+\dots~, \nonumber  \\
\eeqn
where here c.c. indicate complex conjugation of entire lines. Several terms in the  Lagrangian \eqref{SDBIgammathetawithsolvedD} still remain to be eliminated. 

\paragraph{\bf Eliminating dimension-8 and dimension-10 terms.}
The dimension-10 operators in the last line of \eqref{SDBIgammathetawithsolvedD} can be eliminated through the field redefinition 
\beqn
 &&\lambda_{\alpha}\rightarrow\lambda_{\alpha}+h (\sigma^{\mu\nu}\lambda)_{\alpha}\left(F_{\mu\nu}F^2- F_{\mu\nu}F\Ft \right), \qquad h=-i \frac{4\kappa^4 d_0^2}{\sfD_0(1+8\kappa^2\gamma^2+\sqrt{Z_0})}, \label{fieldrederfdim10thetagammadirect}
\eeqn
which is the analog to \eqref{fieldrederfdim10} used in \cref{subsecphysicalNL}. The fermion kinetic terms transform as in \eqref{gauginokineticunderfieldredef10} and cancel the dimension-10 operators.
 
Dimension-8 terms in the fourth and fifth line of \eqref{SDBIgammathetawithsolvedD} can also be eliminated by field redefinitions
\beqn\label{eliminatedim8}
&&\ld_{\alpha}\rightarrow\ld_{\alpha}+b\ld_{\alpha} F^2, \qquad \ld_{\alpha}\rightarrow\ld_{\alpha}+c\ld_{\alpha} F\Ft, \label{fieldrederfdim8thetagammadirect}
\eeqn
with appropriate $b,c$ coefficients. This is again analog to the field redefinitions \eqref{fieldredef6} and  \eqref{fieldredef7} in \cref{subsecphysicalNL}. Acting \eqref{eliminatedim8} on fermion kinetic terms eliminates the above dimension-8 operators. 
Additional terms generated by the field redefinitions \eqref{fieldrederfdim10thetagammadirect} and \eqref{fieldrederfdim8thetagammadirect} have dimension at least 12.
The leftover dimension-8 operators containing two gauginos and two gauge bosons are those of the third line and correspond to the standard goldstino coupling to the energy momentum tensor, foretold by the low energy theorems.

Therefore, we can discard the last three lines of~\eqref{SDBIgammathetawithsolvedD} by using  \eqref{fieldrederfdim10thetagammadirect} and \eqref{fieldrederfdim8thetagammadirect}, and   the physical on-shell Lagrangian contains only the first three lines of \eqref{SDBIgammathetawithsolvedD}.


 \paragraph{Rescaling and final result.}
Finally we rescale the fields to obtain canonical kinetic terms 
\beqn
\ld\ldb &\rightarrow&  \frac{\sGo \left(1+8\kappa^2\gamma^2+\sGo \right)}{2(1+8\kappa^2\gamma^2)}\ld\ldb ~, \\
F_{\mu\nu } &\to&   Z_0^{1/4 } F_{\mu\nu } ~.
 \label{fieldrenormthetagamma}
\eeqn
 Applying this to the first three lines of \eqref{SDBIgammathetawithsolvedD}, we   finally arrive at the  following   on-shell Lagrangian  \beqn\label{finalLwithgammatheta}
\mathcal{L}_{{\rm SDBI}+\gamma,\theta}
&=& \frac{1}{8 g^2 \kappa ^2}\left(1-\frac{ \kappa ^2}{\bar{\kappa}^2}\right) + \frac{i\theta F\Ft\bar{\kappa}^2}{32 \pi^2 \kappa^2}-\frac {F^2}{4 g^2} +\frac{\bar{\kappa}^2}{4g^2}F^4-\frac{\bar{\kappa}^2}{4g^2}\left( F\Ft\right)^2 \nonumber \\[5pt]
&&- \frac{1}{2g^2}\left(i\ldbl + {\rm c.c.}\right)- \frac{\bar{\kappa}^2}{g^2}\p_{\mu}(\ld^2)\p^{\mu}(\ldb^2)+ \frac{2\bar{\kappa}^2}{g^2} (i \lambda \sigma^{\rho}\partial_{\nu} \bar{\lambda} -i \partial_{\nu} \lambda \sigma^{\rho} \bar{\lambda})F^{\nu}_{~\mu} F^{\mu}_{~\rho} \qquad\nonumber\\[5pt]
&&+~O({\rm dim} ~ 12), 
\eeqn
where we defined 
\beqn \label{kappathetagamma}
\bar{\kappa}^2=\kappa^2\frac{\sGo}{1+8\kappa^2\gamma^2}
=\frac{2\sqrt2 \pi^2   \kappa^2}{\sqrt{8\pi^4(1+8\kappa^2\gamma^2) +g^4 \gamma^2 \kappa^2 \theta^2}}\,.
 \eeqn
 This perturbative low energy expansion agrees with \eqref{Lagrangianformdim12NL6onshell}, up to an additive constant which plays no role in global supersymmetry. 

After dropping the total derivative term $\theta F\tilde F$, it is easy to see that this expansion agrees on-shell with the low energy expansion of the action 
\be\label{DBIconstarint}
\mathcal L'_{{\rm SDBI}+\gamma,\theta}= \frac{1}{8 g^2 \kappa ^2}\left(1+\frac{ \kappa ^2}{\bar{\kappa}^2}\right)  -\frac{1}{8g^2 \bar \kappa^2}\det{\bm A} \Big(1+ \sqrt{-\det (\eta_{\mu\nu} +2\sqrt2 \bar\kappa \mathcal F_{\mu\nu})} \Big)\,.
\ee
 One can also compare this action with the bosonic truncation given in \eqref{DBIbosonic} which can be rewritten as follows
 \beqn\label{DBIboson2}
 \mathcal L_{bosonic} &  =&\frac{1 }{8g^2 \kappa^2}
   -\frac {1}{8 g^2 \bar \kappa^2 }  
   \sqrt{-  \det\Big(\eta_{\mu\nu}+2 \sqrt{2}  \bar \kappa F_{\mu\nu}\Big) }  ~. \qquad
 \eeqn 
 where we have rescaled $F$ as $  F_{\mu\nu } \to   Z_0^{1/4 } F_{\mu\nu } $ and dropped the total derivative term $\theta F \tilde F$. It is obvious that \eqref{DBIboson2} indeed agrees with the bosonic truncation of \eqref{DBIconstarint} by setting  $\lambda=0$ and thus $\det{\bm A}=1$.  
Instead, in the pure fermionic case $F=0$,  \eqref{DBIconstarint} becomes the VA action, in agreement with the well-known fact that the VA action is the low energy description of spontaneous supersymmetry breaking. 
 Considering our explicit low energy expansion up to dimension 10 as well as the above limits, we conclude that \eqref{DBIconstarint} is indeed  on-shell equivalent to the original SDBI$+ {\gamma,\theta}$ action. 
 
 To study the SDBI+FI model, we can consider the double scaling limit $\gamma \rightarrow 0$ with  $\gamma\theta =-{8\sqrt{2}\pi^2}\xi $ fixed, as explained in \eqref{limitcases}.  In this limit, the value of $\bar\kappa$ in  \eqref{kappathetagamma} gives the value in \eqref{constantsdefinedhere}. Hence the result \eqref{DBIconstarint}
 also agrees with the explicit computation~\eqref{nonlinLagrangianfinal}  in the last section based on the non-linear formalism. Therefore, this also provides a non-trivial test of the non-linear supersymmetric formalism of~\cite{Cribiori:2018dlc}.

\section{Summary and outlook}\label{Summary}

In this work, we have studied the on-shell SDBI action implemented with either a standard FI term or an induced FI term through a $\gamma$ supersymmetry deformation in the presence of a $\theta$-angle. We have computed its low-energy expansion up to mass dimension-12 terms. 
We argued that the result up to dimension-8 operators can be guessed by non-linear supersymmetry and its low energy theorem, and thus that the first non-trivial computation starts for operators of dimension 10. We have shown that these operators vanish on-shell and can be eliminated by field redefinitions, while the operators of dimension 12 involving up to two gauginos are reduced to those dictated by the low-energy theorem.

Our result suggests that in either case, the deformation or the FI parameter  only renormalize  the couplings (without changing the form) of the physical on-shell standard SDBI action. Based on the bosonic truncation, it was argued that  the deformation or the FI parameters in SDBI action do not break the supersymmetry completely; instead they rotate the remaining residual supersymmetry. Considering the nature of the SDBI action realizing partial supersymmetry breaking with both linear and non-linear supersymmetry,  it is not surprising to see the trivial role of the deformation or the FI parameter on-shell. On the other hand, the rotation modifies the field transformations of the linear supersymmetry in a non-linear way (although without constant in the gaugino transformation) that makes the result non-trivial.

Obviously, the rotation argument breaks down in the presence of another (referent) SDBI action and supersymmetry breaking should occur in this system. An interesting question remains, whether there is a deformation of the SDBI action that breaks spontaneously the linear supersymmetry and its coupling to supergravity. A related question concerns the effective field theory of branes in string (or M) theory in the presence or not of supersymmetry breaking.

 \section*{Acknowledgments}
 This work was supported in part by the Swiss National Science Foundation, in part by the Labex
``Institut Lagrange de Paris" and in part by a CNRS PICS grant. We would like to thank Jean-Pierre Derendinger  for discussions.
 

 \appendix 
 \section{Conventions and useful identities}\label{usefulids}
 We use conventions of \cite{wess1992supersymmetry} for spinors.
 
We list some properties of Pauli $\sigma$-matrices as follows:
\beqn
&& \hspace{-1cm}\sigma^{\mu\nu}\equiv\frac{1}{4}(\sigma^{\mu}\bar{\sigma}^{\nu}- \sigma^{\nu}\bar{\sigma}^{\mu}),  \hspace{1cm}  \sigma^{\mu}\bar{\sigma}^{\nu}+ \sigma^{\nu}\bar{\sigma}^{\mu} = -2 \eta^{\mu\nu}  \hspace{0.5cm} \rightarrow  \hspace{0.5cm}   \sigma^{\mu}\bar{\sigma}^{\nu}=-\eta^{\mu\nu}+2\sigma^{\mu\nu}, \\
&& \hspace{-1cm}{\rm Tr}(\sigma^{\mu\nu}\sigma^{\rho\gamma})=-\frac12(\eta^{\mu\rho}\eta^{\nu\gamma}-\eta^{\mu\gamma}\eta^{\nu\rho})-\frac i2\epsilon^{\mu\nu\rho\gamma},  \\
&& \label{gammamatrices} \hspace{-1cm}\sigma^{\mu\nu}{\sigma}^{\rho}=\frac 12 \left(\eta^{\mu\rho}\sigma^{\nu}-\eta^{\nu\rho}\sigma^{\mu}+i \epsilon^{\mu\nu\rho\gamma}\sigma_{\gamma}\right),  \hspace{0.7cm}\bar{\sigma}^{\mu}\sigma^{\nu\rho}=-\frac 12 \left(\eta^{\mu\nu}\bar{\sigma}^{\rho}-\eta^{\mu\rho}\bar{\sigma}^{\nu}+i \epsilon^{\mu\nu\rho\gamma}\bar{\sigma}_{\gamma} \right).
\eeqn

We define the dual $\Ft^{\gamma \rho}$ of the antisymmetric field-strength tensor $F_{\mu\nu}$ and the associated self-dual or anti-self dual tensors as follows
\beqn \label{dualtensors}
\Ft^{\gamma\rho}=\frac i2 \epsilon^{\gamma\rho\mu\nu}F_{\mu\nu}, \qquad F^+_{\mu\nu}=\frac{F_{\mu\nu}+\Ft_{\mu\nu}}2,  \qquad F^-_{\mu\nu}=\frac{F_{\mu\nu}-\Ft_{\mu\nu}}2.
\eeqn
The above tensors satisfy the following properties
\beqn
&&\Ft^{\mu}_{~\,\rho}\Ft^{\rho\nu}=-\frac 12 \eta^{\mu\nu}F^2-F^{\mu}_{~\rho}F^{\rho\nu}~, \qquad F^2\equiv F^{\mu\nu}F_{\mu\nu}=\Ft^{2} ~, \\
&&F^{\mu\rho}F_{\rho\alpha}F^{\alpha\nu}=\frac 14 \Ft^{\mu\nu}F\Ft-\frac 12 F^{\mu\nu}F^2, \qquad  F_{\mu\alpha}\Ft^{\alpha}_{~\,\nu}=\frac 14 \eta_{\mu\nu}F\Ft~, \\
&&F^{+\mu}_{~~~\rho}F^{-\rho\nu}=F^{+\nu}_{~~~\rho}F^{-\rho\mu}=\frac 14\left(F^{\mu}_{~\rho}F^{\rho\nu}-\Ft^{\mu}_{~\,\rho}\Ft^{\rho\nu}\right)=\frac 18 \eta_{\mu\nu} F^2 + \frac 12 F^{\mu}_{~\rho}F^{\rho\nu} .
\eeqn
From \eqref{gammamatrices} and \eqref{dualtensors} we derive the useful identities 
\beqn
\sigma^{\mu}\bar{\sigma}^{\nu\rho}F_{\nu\rho}=-2F^{-\mu}_{~~~\nu}\sigma^{\nu}, \qquad F_{\nu\rho}\sigma^{\nu\rho}\sigma^{\mu}=2F^{+\mu}_{~~~\nu}\sigma^{\nu}.
\eeqn

We also present here useful spinor conventions and algebra used for our calculation:
\begin{flalign}
&~~~\psi\chi=\psi^{\alpha}\chi_{\alpha}=-\chi_{\alpha}\psi^{\alpha}=\chi^{\alpha}\psi_{\alpha}=\chi\psi,\hspace{0.7cm}\psi^{\alpha}=\epsilon^{\alpha\beta}\psi_{\beta},&\\[5pt]
&~~~ \chi\sigma^{\mu}\bar{\psi}=-\bar{\psi}\bar{\sigma}^{\mu}\chi,\hspace{0.7cm} \left(\chi\sigma^{\mu}\bar{\psi}\right)^{*}=\psi\sigma^{\mu}\bar{\chi},\hspace{0.7cm} \chi\sigma^{\mu\nu}\psi=-\psi\sigma^{\mu\nu}\chi, \hspace{0.7cm} \left(\chi\sigma^{\mu\nu}\psi \right)^{*}=\bar{\chi}\bar{\sigma}^{\mu\nu}\bar{\psi},\hspace{-0.1cm}&\\[5pt]
&~~~\theta_{\alpha}\theta_{\beta}=\frac 12\epsilon_{\alpha\beta}\theta\theta,\hspace{0.7cm}\theta^{\alpha}\theta^{\beta}=-\frac 12\epsilon^{\alpha\beta}\theta\theta,\hspace{0.7cm}
\bar{\theta}_{\dot{\alpha}}\bar{\theta}_{\dot{\beta}}=-\frac 12\epsilon_{\dot{\alpha}\dot{\beta}}\bar{\theta}\bar{\theta},\hspace{0.7cm}\bar{\theta}^{\dot{\alpha}}\bar{\theta}^{\dot{\beta}}=\frac 12\epsilon^{\dot{\alpha}\dot{\beta}}\bar{\theta}\bar{\theta}~,& \\[5pt]
&~~~\chi_{\alpha}\equiv(\sigma^{\mu\nu}\lambda)_{\alpha} \implies  \chi^{\alpha}=-(\lambda\sigma^{\mu\nu})^{\alpha}, ~~~\bar{\chi}^{\dot{\alpha}}=(\bar{\sigma}^{\mu\nu}\bar{\lambda})^{\dot{\alpha}}, ~~~\bar{\chi}_{\dot{\alpha}}=-(\bar{\lambda}\bar{\sigma}^{\mu\nu})_{\dot{\alpha}} ~,&\\[5pt]
&~~~\theta\psi \, \theta\phi=-\frac12 \theta\theta \, \psi\phi,\hspace{0.7cm}\psi\chi \,\bar{\varphi}\bar{\eta}=-\frac12~ \psi\sigma^{\mu}\bar{\varphi} ~\chi\sigma_{\mu}\bar{\eta} \quad{\rm (Fierz ~ identity)}~. \label{Fierzid}&
\end{flalign}

 \section{Some computational details}\label{somedetails}

\subsection{Derivation of \eqref{Lnonlintsf}} \label{derivationOfgaugeInv}
The extra term in the second line of \eqref{Lnonlin}, containing the gauge potential $u_\mu $, arises from the FI term. To make gauge invariance manifest, we rewrite it in terms of the field-strength $F_{\mu \nu}$ as we describe below. 
Using the following property of determinant
 \beqn
 \epsilon^{abcd} ({\bm A}^{-1})_{a}^{~\nu}({\bm A}^{-1})_{c}^{~\rho}({\bm A}^{-1})_d^{~\mu}({\bm A}^{-1})_b^{~\gamma}=\epsilon^{\nu \gamma \rho \mu} \det{\bm A}^{-1},
 \eeqn
the second line of \eqref{Lnonlin} can be written as 
\beqn \label{relationepsilontensordetA}
i \det{\bm A} \epsilon^{abcd}[({\bm A}^{-1})_{a}^{~\nu}\partial_{\nu} \lambda]\sigma_b[({\bm A}^{-1})_{c}^{~\rho}\partial_{\rho}\bar{\lambda}]({\bm A}^{-1})_d^{~\mu}u_{\mu} 
&=&i \det{\bm A} ~ \epsilon^{\nu \gamma \rho \mu} ~\det{\bm A}^{-1} {\bm A}^{e}_{\gamma} ~  \partial_{\nu} \lambda \sigma_e \partial_{\rho}\bar{\lambda} u_{\mu} \nonumber \\
&=&i \epsilon^{\nu \gamma \rho \mu} {\bm A}^{e}_{\gamma} ~  \partial_{\nu} \lambda \sigma_e \partial_{\rho}\bar{\lambda} u_{\mu}~ .
\eeqn
This can be further simplified by using several   integrations  by parts and Fierz identities~\eqref{Fierzid}
\beqn \label{simplifiedFicontri}
i \epsilon^{\nu \gamma \rho \mu} {\bm A}^{e}_{\gamma} ~  \partial_{\nu} \lambda \sigma_e \partial_{\rho}\bar{\lambda} u_{\mu} &=& i \epsilon^{\nu \gamma \rho \mu} (\delta^{e}_{\gamma}+i\kappa^2\ld\sigma^e\p_{\gamma}\ldb -i\kappa^2\p_{\gamma}\ld \sigma^e\ldb) ~  \partial_{\nu} \lambda \sigma_e \partial_{\rho}\bar{\lambda} u_{\mu} \nonumber \\
&=&-  i \epsilon^{\nu \gamma \rho \mu}  \delta^{e}_{\gamma} ~ \lambda \sigma_e \partial_{\rho}\bar{\lambda} ~ \partial_{\nu}u_{\mu} - i \kappa^2 \epsilon^{\nu \gamma \rho \mu}(i\ld\sigma^e\p_{\gamma}\ldb -i\p_{\gamma}\ld \sigma^e\ldb) ~ \lambda \sigma_e \partial_{\rho}\bar{\lambda}~ \partial_{\nu} u_{\mu} 
\nonumber \\[5pt]
&&- i\kappa^2 \epsilon^{\nu \gamma \rho \mu} \partial_{\nu}(i\ld\sigma^e\p_{\rho}\ldb -i\p_{\rho}\ld \sigma^e\ldb) ~ \lambda \sigma_e \partial_{\rho}\bar{\lambda} ~ u_{\mu}   +\text{total   derivative}
 \nonumber \\
&=& -  \frac i2 \epsilon^{ \gamma \rho \nu \mu} ~ \lambda \sigma_{\gamma} \partial_{\rho}\bar{\lambda} ~ F_{\nu \mu} -\frac i2 \kappa^2\epsilon^{\nu\gamma\rho\mu} (-2) (i \ld^2 ~ \p_{\gamma}\ldb\p_{\rho}\ldb-i\ld\p_{\gamma}\ld ~\ldb \p_{\rho}\ldb)F_{\nu\mu} \nonumber \\
&&-i \kappa^2\epsilon^{\nu\gamma\rho\mu} (-2) (i\ld \p_{\nu} \ld \p_{\gamma} \ldb \p_{\rho}\ldb - i \ld \p_{\gamma}\ld \p_{\nu}\ldb\p_{\rho}\ldb)~u_{\mu}+\text{total   derivative} \nonumber \\
&=& -  \lambda \sigma_{\gamma} \partial_{\rho}\bar{\lambda} ~ \Ft^{\gamma \rho} +0 + \frac i4\kappa^2 \epsilon^{\nu\gamma\rho\mu} \p_{\gamma}(\ld^2)\p_{\rho}(\ldb^2)F_{\nu\mu} + 0+\text{total   derivative} \nonumber \\
&=&-  \lambda \sigma_{\gamma} \partial_{\rho}\bar{\lambda} ~ \Ft^{ \gamma \rho} -\frac i4\kappa^2 \epsilon^{\nu\gamma\rho\mu} \ld^2\p_{\rho}(\ldb^2)\p_{\gamma}F_{\nu\mu} +\text{total   derivative} \nonumber \\
&=& -  \lambda \sigma_{\gamma} \partial_{\rho}\bar{\lambda} ~ \Ft^{ \gamma \rho}+\text{total   derivative} .
\eeqn
We also repeatedly used relations
$\epsilon^{\nu\gamma\rho\mu} \p_\mu \p_\nu  =0  $ and $\epsilon^{\nu\gamma\rho\mu} \p_\mu \lambda \p_\nu\lambda =0$. 
Once rewritten as \eqref{simplifiedFicontri} it is obvious that the second line of \eqref{Lnonlin}  is  gauge invariant.

 \subsection{Superfield expansions}\label{superfieldexpansion}
 
We first recall the component expansions of the superfields  $W_\alpha$ and $W^2$  
\beqn \label{definitionnotationsfielddirectcalcl}
&&W_\alpha = -i \lambda_\alpha +  \theta_\alpha \sfD  -  i (\sigma^{\mu\nu}\theta)_\alpha F_{\mu\nu}    +\taa (\sigma^\mu   \p_\mu {\bar \lambda}  )_\alpha~, \\[5pt]
&& W^2= C+2\psi\ta +\taa E~, \\ [3pt]
&& C=-\lambda^2, \quad \psi_{\beta}=F_{\mu\nu} (\sigma^{\mu\nu} \lambda)_\beta -i \sfD \lambda_\beta \equiv \Psi_\beta-i \sfD \lambda_\beta, \quad E=\sfD^2-\frac12 (F^2+ F \Ft)-2i \ld \slashed{\p} \ldb~.  \label{defofPsi} 
 \hspace{1cm} 
\eeqn 
We also use the following chiral and anti-chiral superfield expansions
\beqn 
-\frac 14 \overline{D}^2\overline{W}^2(y)=\bar{E}+2i\ta\slashed{\p}\bar{\psi}+\ta^2\Box\bar{C}~, \label{D2W2}\\
-\frac 14 {D}^2{W}^2(\bar{y})={E}+2i\tab\bar{\slashed{\p}}{\psi}+\ta^2\Box{C}~.\label{Db2Wb2}
\eeqn
Then, the chiral superfield $\Phi$  defined in \eqref{superfieldsdefinition} has the following field component expansion, depending of the chiral coordinates $y^\mu=x^\mu+i\ta\sigma^\mu\tab$,
\beqn
&&\Phi(y)= \frac{W^2}{\vphantom{{1}^2} \overline{D}^2\overline{W}^2}(y)= \phi (y)+ \chi(y) \theta + \theta^2 \mathcal{G}(y), \\
&&\phi=-\frac{C}{4 \bar{E}}, \qquad \chi_{\alpha}=-\frac{\psi_{\alpha}}{2 \bar{E}}+i \frac{ C (\slashed{\partial}\bar{\psi})_{\alpha}}{2 \bar{E}^2}, \qquad \mathcal{G}=-\frac{E}{4 \bar{E}}-\frac{i\psi\slashed{\partial}\bar{\psi}}{2\bar{E}^2}+\frac{C\Box \bar{C}}{4 \bar{E}^2}-\frac{C (\slashed{\partial}\bar{\psi})^2}{2 \bar{E}^3}~.\qquad
\eeqn
We can now compute the component expansion of  the real superfield $\Phi \bar{\Phi}$  
\beqn
\Phi \bar{\Phi}(x)&=& \phi \bar{\phi} +( \bar{\chi} \bar{\theta} \phi + {\rm c.c.} ) + (\tab^2 {\phi}\bar{\mathcal{G}} + {\rm c.c.}) + \chi \theta \bar{\chi} \bar{\theta} + i \theta \sigma^{\mu} \bar{\theta} \left(\bar{\phi}\partial_{\mu}\phi-\phi \partial_{\mu}\bar{\phi}\right) \nonumber\\[5pt]
&& + \left(\bar{\theta}^2\theta \left[ \bar{\mathcal{G}}\chi-\frac i2 \sigma^{\mu}\left(\bar{\chi}\partial_{\mu}\phi-\partial_{\mu}\bar{\chi}\phi\right)\right] + {\rm c.c.}\right)  \\
&&+ ~\theta^2\bar{\theta}^2 \left[ \mathcal{G}\bar{\mathcal{G}} - \partial^{\mu} \phi \partial_{\mu} \bar{\phi} -\frac i4 \chi \slashed{\partial}\bar{\chi} + \frac i4 \partial_{\mu}\chi \sigma^{\mu} \bar{\chi}+ \frac 14 \Box(\phi\bar{\phi})\right] . \nonumber
\eeqn

Finally, for the real superfields $\mathcal{A}$ and $\mathcal{B}$  defined in \eqref{AtBt}, we have the following  component expansions
\beqn
\mathcal{A}(x)&=&\frac{\kappa^2}{2} (D^2 W^2+\overline D^2 \overline W^2 )\nonumber\\
&=&-2\kappa^2\left[\vphantom{\frac 14}\left(\sfD^2+\bar{\sfD}^2\right)+\mathcal{E}_+ + 2i\left(\bar{\theta}\bar{\sigma}^{\mu}\partial_{\mu}(\Psi-i\sfD\lambda) + \theta \sigma^{\mu}\partial_{\mu}(\bar{\Psi}+i\bar{\sfD}\bar{\lambda})\right)\right.\nonumber \\
&&\hspace{40pt}\left.  +~ \bar{\theta}\bar{\theta}\Box C+\theta \theta \Box \bar{C}  -i\theta\sigma^{\mu}\bar{\theta}\partial_{\mu}(\mathcal{E}_-+\sfD^2-\Db^2)\right.\nonumber \\
&&\hspace{40pt}\left.+~ \bar{\theta}\bar{\theta}\theta \Box(\Psi-i\sfD\lambda)+\theta \theta \bar{\theta} \Box(\bar{\Psi}+i\bar{\sfD}\bar{\lambda})+\theta^2\bar{\theta}^2\frac 14 \Box \left(\mathcal{E}_+ + \sfD^2+\bar{\sfD}^2\right) \right],
\\[5pt]
\mathcal{B}(x)&=&i\frac{\kappa^2}{2} (D^2 W^2-\overline D^2 \overline W^2 )\nonumber\\
&=&-2i\kappa^2\left[\vphantom{\frac 14}\left(\sfD^2-\bar{\sfD}^2\right)+\mathcal{E}_- + 2i\left(\bar{\theta}\bar{\sigma}^{\mu}\partial_{\mu}(\Psi-i\sfD\lambda) - \theta \sigma^{\mu}\partial_{\mu}(\bar{\Psi}+i\bar{\sfD}\bar{\lambda})\right) \right.\nonumber \\
&&\hspace{40pt}\left.+~\bar{\theta}\bar{\theta}\Box C-\theta \theta \Box \bar{C}  -i\theta\sigma^{\mu}\bar{\theta}\partial_{\mu}(\mathcal{E}_++\sfD^2+\Db^2)\right.\nonumber\\
&&\hspace{40pt}\left.+~\bar{\theta}\bar{\theta}\theta \Box(\Psi-i\sfD\lambda)-\theta \theta \bar{\theta} \Box(\bar{\Psi}+i\bar{\sfD}\bar{\lambda})+\theta^2\bar{\theta}^2\frac 14 \Box \left(\mathcal{E}_- + \sfD^2-\bar{\sfD}^2\right) \right],
\eeqn
where
 \beqn \label{defEpm}
 \mathcal{E}_{\pm} \equiv E \pm \bar{E}- \left(\sfD^2 \pm \bar{\sfD}^2\right).
 \eeqn
 Since the auxiliary fields  $\sfD$ and $\Db$  are not dynamical and should be eliminated at the end, we   can isolate their contribution in the above two real superfields
\beqn
&\mathcal{A}&\equiv\mathcal{A}_{scalar}+\mathcal{A}'=-2\kappa^2\left(\sfD^2+\bar{\sfD}^2\right)+\mathcal{A}'=-4\kappa^2\left(d^2-\gamma^2\right) +\mathcal{A}', \\
&\mathcal{B}&\equiv\mathcal{B}_{scalar}+\mathcal{B}'=-2i\kappa^2\left(\sfD^2-\bar{\sfD}^2\right)+\mathcal{B}'=8\kappa^2d\gamma +\mathcal{B}',
\eeqn
Then, the superfield $\mathcal{M}$ defined in \eqref{superfieldsdefinition} can be expanded up to mass dimension 10 (included)
 \beqn
\M&=&  1+ \mathcal A - \sqrt{1+2 \mathcal A- \mathcal B^2} \nonumber\\
&=&1-4\kappa^2\left(d^2-\gamma^2\right) +\mathcal{A}' -\sqrt{1-8\kappa^2\left(d^2-\gamma^2\right)-64\kappa^4d^2\gamma^2 +2\mathcal{A}'-16\kappa^2d\gamma \mathcal{B}' -{\mathcal{B}'}^2} \nonumber \\
&=&1-4\kappa^2\left(d^2-\gamma^2\right) +\mathcal{A}' \nonumber -\sqrt{Z}\left[1+ \frac {\cA'}{Z}-\frac{8\kappa^2 d\gamma \cB'}{Z}-\frac{{\cB'}^2}{2Z} \right.\nonumber \\
&& \left. -\frac 1{8Z^2}\left(2\cA'-{\cB'}^2-16\kappa^2d\gamma\cB'\right)^2+\frac{1}{16Z^3}\left(2\cA'-16\kappa^2d\gamma\cB'\right)^3\right] + O(\kappa^6).
\eeqn
where we introduced 
\beqn\label{labelZ}
Z =(1+8\kappa^2\gamma^2)(1-8\kappa^2d^2)=1-8\kappa^2\left(d^2-\gamma^2\right)-(8\kappa^2d\gamma)^2~.
\eeqn

From \eqref{defofPsi} we can compute explicitly the useful expansion
\beqn \label{psidpsi}
i\psi\slashed{\p}\bar{\psi}&=&i\Psi\slashed{\p}\bar{\Psi}-\Db\Psi\slp\ldb-\p_{\mu}(\Db)\Psi\sigma^{\mu}\ldb+\sfD \ld \slp \bar{\Psi} + i\sfD\Db\ldbl \nonumber \\
&=&2i\ld\sigma^{\mu}\p_{\nu}\ldb F_{\mu}^{~\rho}F_{\rho}^{~\nu}+\frac i2 \ldbl F^2+\Db\ld\sigma^{\mu}\p_{\nu}\ldb(F^{\nu}_{~\mu}+F^{*\nu}_{~~\mu})+\p^{\nu}(\Db)\lsl(F_{\nu\mu}+\Ft_{~\nu\mu})\nonumber\\
&&\quad - \sfD\ld\sigma^{\mu}\p^{\nu}\ldb(F_{\nu\mu}-\Ft_{~\nu\mu})-\sfD\lsl\p^{\nu}F_{\nu\mu}+i\sfD\Db\ldbl + i \sfD \p_{\mu}\Db \lsl~.
\eeqn


  \bibliography{Note_SDBI+FI}

\providecommand{\href}[2]{#2}\begingroup\raggedright\begin{thebibliography}{10}

\bibitem{Bagger:1996wp}
J.~Bagger and A.~Galperin, \emph{{A New Goldstone multiplet for partially
  broken supersymmetry}},
  \href{http://dx.doi.org/10.1103/PhysRevD.55.1091}{\emph{Phys. Rev.} {\bf D55}
  (1997) 1091--1098}, [\href{https://arxiv.org/abs/hep-th/9608177}{{\tt
  hep-th/9608177}}].

\bibitem{Rocek:1997hi}
M.~Rocek and A.~A. Tseytlin, \emph{{Partial breaking of global D = 4
  supersymmetry, constrained superfields, and three-brane actions}},
  \href{http://dx.doi.org/10.1103/PhysRevD.59.106001}{\emph{Phys. Rev.} {\bf
  D59} (1999) 106001}, [\href{https://arxiv.org/abs/hep-th/9811232}{{\tt
  hep-th/9811232}}].

\bibitem{Antoniadis:2008uk}
I.~Antoniadis, J.~P. Derendinger and T.~Maillard, \emph{{Nonlinear N=2
  Supersymmetry, Effective Actions and Moduli Stabilization}},
  \href{http://dx.doi.org/10.1016/j.nuclphysb.2008.09.008}{\emph{Nucl. Phys.}
  {\bf B808} (2009) 53--79}, [\href{https://arxiv.org/abs/0804.1738}{{\tt
  0804.1738}}].

\bibitem{Antoniadis:2017jsk}
I.~Antoniadis, J.-P. Derendinger and C.~Markou, \emph{{Nonlinear $
  \mathcal{N}=2 $ global supersymmetry}},
  \href{http://dx.doi.org/10.1007/JHEP06(2017)052}{\emph{JHEP} {\bf 06} (2017)
  052}, [\href{https://arxiv.org/abs/1703.08806}{{\tt 1703.08806}}].

\bibitem{Antoniadis:2019gbd}
I.~Antoniadis, H.~Jiang and O.~Lacombe, \emph{{$\mathcal{N} = 2$ supersymmetry
  deformations, electromagnetic duality and Dirac-Born-Infeld actions}},
  \href{http://dx.doi.org/10.1007/JHEP07(2019)147}{\emph{JHEP} {\bf 2019} (Jul,
  2019) 147}.

\bibitem{Bellucci:2015qpa}
S.~Bellucci, N.~Kozyrev, S.~Krivonos and A.~Sutulin, \emph{{Space-filling
  D3-brane within coset approach}},
  \href{http://dx.doi.org/10.1007/JHEP08(2015)094}{\emph{JHEP} {\bf 08} (2015)
  094}, [\href{https://arxiv.org/abs/1505.07386}{{\tt 1505.07386}}].

\bibitem{Cribiori:2018dlc}
N.~Cribiori, F.~Farakos and M.~Tournoy, \emph{{Supersymmetric Born-Infeld
  actions and new Fayet-Iliopoulos terms}},
  \href{http://dx.doi.org/10.1007/JHEP03(2019)050}{\emph{JHEP} {\bf 03} (2019)
  050}, [\href{https://arxiv.org/abs/1811.08424}{{\tt 1811.08424}}].

\bibitem{Volkov:1972jx}
D.~V. Volkov and V.~P. Akulov, \emph{{Possible universal neutrino
  interaction}}, {\emph{JETP Lett.} {\bf 16} (1972) 438--440}.

\bibitem{Kuzenko:2010ef}
S.~M. Kuzenko and S.~J. Tyler, \emph{{Relating the Komargodski-Seiberg and
  Akulov-Volkov actions: Exact nonlinear field redefinition}},
  \href{http://dx.doi.org/10.1016/j.physletb.2011.03.020}{\emph{Phys. Lett.}
  {\bf B698} (2011) 319--322}, [\href{https://arxiv.org/abs/1009.3298}{{\tt
  1009.3298}}].

\bibitem{Tseytlin:1999dj}
A.~A. Tseytlin, \emph{{Born-Infeld action, supersymmetry and string theory}},
  \href{http://dx.doi.org/10.1142/9789812793850_0025}{\emph{Shifman, M.A.
  (ed.): The many faces of the superworld} (1999) 417--452},
  [\href{https://arxiv.org/abs/hep-th/9908105}{{\tt hep-th/9908105}}].

\bibitem{Bergshoeff:2001dc}
E.~A. Bergshoeff, A.~Bilal, M.~de~Roo and A.~Sevrin, \emph{{Supersymmetric
  nonAbelian Born-Infeld revisited}},
  \href{http://dx.doi.org/10.1088/1126-6708/2001/07/029}{\emph{JHEP} {\bf 07}
  (2001) 029}, [\href{https://arxiv.org/abs/hep-th/0105274}{{\tt
  hep-th/0105274}}].

\bibitem{Antoniadis:2004uk}
I.~Antoniadis and M.~Tuckmantel, \emph{{Nonlinear supersymmetry and
  intersecting D-branes}},
  \href{http://dx.doi.org/10.1016/j.nuclphysb.2004.07.027}{\emph{Nucl. Phys.}
  {\bf B697} (2004) 3--47}, [\href{https://arxiv.org/abs/hep-th/0406010}{{\tt
  hep-th/0406010}}].

\bibitem{Samuel:1982uh}
S.~Samuel and J.~Wess, \emph{{A Superfield Formulation of the Nonlinear
  Realization of Supersymmetry and Its Coupling to Supergravity}},
  \href{http://dx.doi.org/10.1016/0550-3213(83)90622-3}{\emph{Nucl. Phys.} {\bf
  B221} (1983) 153--177}.

\bibitem{wess1992supersymmetry}
J.~Wess and J.~Bagger, \emph{Supersymmetry and supergravity}.
\newblock Princeton university press, 1992.

\bibitem{Cribiori:2017laj}
N.~Cribiori, F.~Farakos, M.~Tournoy and A.~van Proeyen, \emph{{Fayet-Iliopoulos
  terms in supergravity without gauged R-symmetry}},
  \href{http://dx.doi.org/10.1007/JHEP04(2018)032}{\emph{JHEP} {\bf 04} (2018)
  032}, [\href{https://arxiv.org/abs/1712.08601}{{\tt 1712.08601}}].

\bibitem{Antoniadis:2018cpq}
I.~Antoniadis, A.~Chatrabhuti, H.~Isono and R.~Knoops, \emph{{Fayet--Iliopoulos
  terms in supergravity and D-term inflation}},
  \href{http://dx.doi.org/10.1140/epjc/s10052-018-5861-6}{\emph{Eur. Phys. J.}
  {\bf C78} (2018) 366}, [\href{https://arxiv.org/abs/1803.03817}{{\tt
  1803.03817}}].

\bibitem{Antoniadis:2018oeh}
I.~Antoniadis, A.~Chatrabhuti, H.~Isono and R.~Knoops, \emph{{The cosmological
  constant in Supergravity}},
  \href{http://dx.doi.org/10.1140/epjc/s10052-018-6175-4}{\emph{Eur. Phys. J.}
  {\bf C78} (2018) 718}, [\href{https://arxiv.org/abs/1805.00852}{{\tt
  1805.00852}}].

\bibitem{Antoniadis:2019nwz}
I.~Antoniadis and F.~Rondeau, \emph{{New K{\"a}hler invariant Fayet-Iliopoulos
  terms in supergravity and cosmological applications}},
  \href{https://arxiv.org/abs/1912.08117}{{\tt 1912.08117}}.

\bibitem{Antoniadis:2019hbu}
I.~Antoniadis, J.-P. Derendinger, F.~Farakos and G.~Tartaglino-Mazzucchelli,
  \emph{{New Fayet-Iliopoulos terms in $ \mathcal{N}=2 $ supergravity}},
  \href{http://dx.doi.org/10.1007/JHEP07(2019)061}{\emph{JHEP} {\bf 07} (2019)
  061}, [\href{https://arxiv.org/abs/1905.09125}{{\tt 1905.09125}}].

\bibitem{Kuzenko:2009ym}
S.~M. Kuzenko, \emph{{The Fayet-Iliopoulos term and nonlinear self-duality}},
  \href{http://dx.doi.org/10.1103/PhysRevD.81.085036}{\emph{Phys. Rev.} {\bf
  D81} (2010) 085036}, [\href{https://arxiv.org/abs/0911.5190}{{\tt
  0911.5190}}].

\end{thebibliography}\endgroup
  \bibliographystyle{JHEP} 
  
\end{document}